\documentclass[%
 reprint,
 superscriptaddress,
 amsmath,amssymb,
 aps,
 pra,
]{revtex4-2}

\usepackage{graphicx}
\usepackage{dcolumn}
\usepackage{bm}

\begin{document}

\preprint{APS/123-QED}

\title{Measurements of blackbody radiation-induced transition rates between high-lying $S$, $P$ and $D$ Rydberg levels}


\author{M. Archimi}
\altaffiliation[]{Dipartimento di Fisica “E. Fermi,” Universit\`a di Pisa, Largo Bruno Pontecorvo 3, 56127 Pisa, Italy}
\author{M. Ceccanti}%
\affiliation{Dipartimento di Fisica “E. Fermi,” Universit\`a di Pisa, Largo Bruno Pontecorvo 3, 56127 Pisa, Italy}
\author{M. Distefano}%
\affiliation{Dipartimento di Fisica “E. Fermi,” Universit\`a di Pisa, Largo Bruno Pontecorvo 3, 56127 Pisa, Italy}
\affiliation{INO-CNR, Via G. Moruzzi 1, 56124 Pisa, Italy}
\author{L. Di Virgilio}%
\affiliation{Dipartimento di Fisica “E. Fermi,” Universit\`a di Pisa, Largo Bruno Pontecorvo 3, 56127 Pisa, Italy}
\author{R. Franco}%
\affiliation{Dipartimento di Fisica “E. Fermi,” Universit\`a di Pisa, Largo Bruno Pontecorvo 3, 56127 Pisa, Italy}
\author{A. Greco}%
\affiliation{Dipartimento di Fisica “E. Fermi,” Universit\`a di Pisa, Largo Bruno Pontecorvo 3, 56127 Pisa, Italy}
\author{C. Simonelli}%
\affiliation{Dipartimento di Fisica “E. Fermi,” Universit\`a di Pisa, Largo Bruno Pontecorvo 3, 56127 Pisa, Italy}
\affiliation{INO-CNR, Via G. Moruzzi 1, 56124 Pisa, Italy}
\author{E. Arimondo}%
\affiliation{Dipartimento di Fisica “E. Fermi,” Universit\`a di Pisa, Largo Bruno Pontecorvo 3, 56127 Pisa, Italy}
\affiliation{INO-CNR, Via G. Moruzzi 1, 56124 Pisa, Italy}
\author{D. Ciampini}%
\affiliation{Dipartimento di Fisica “E. Fermi,” Universit\`a di Pisa, Largo Bruno Pontecorvo 3, 56127 Pisa, Italy}
\affiliation{INO-CNR, Via G. Moruzzi 1, 56124 Pisa, Italy}
\author{O. Morsch}%
\affiliation{Dipartimento di Fisica “E. Fermi,” Universit\`a di Pisa, Largo Bruno Pontecorvo 3, 56127 Pisa, Italy}
\affiliation{INO-CNR, Via G. Moruzzi 1, 56124 Pisa, Italy}

\date{\today}

\begin{abstract}
We report experimental measurements of the rates of blackbody radiation-induced transitions between high-lying ($n>60$) S, P and D Rydberg levels of rubidium atoms in a magneto-optical trap using a hybrid field ionization and state-selective depumping technique \cite{PhysRevA.100.030501, PhysRevA.96.043411}. Our results reveal significant deviations of the measured transition rates from theory for well-defined ranges of the principal quantum number. We assume that the most likely cause for those deviations is a modified blackbody spectrum inside the glass cell in which the magneto-optical trap is formed, and we test this assumption by installing electrodes to create an additional microwave cavity around the cell. From the results we conclude that it should be possible to use such external cavities to control and suppress the blackbody radiation-induced transitions. 
\end{abstract}

\maketitle


\section{\label{sec:level1} Introduction}

Rydberg atoms have become a versatile tool for many applications ranging from the detection of electric and magnetic fields \cite{PfauNature2012, PhysRevLett.82.1831, PhysRevLett.59.2947, Fan_2015, Larrouy:19, 6910267} to quantum computation and simulation \cite{RevModPhys.82.2313, ZollerNature2010, PhysRevX.8.011032, BrowaeysNature2020}. In such applications, the strong coupling of Rydberg states to microwave fields can be either a feature or a problem. While the strong coupling is exploited in Rydberg-atom based microwave detectors \cite{TADA2006488, Fan:14, doi:10.1063/1.4997302, PhysRevApplied.15.014053}, the coupling of a Rydberg state to blackbody radiation in the microwave regime leads to transitions to other Rydberg states \cite{PhysRevLett.42.835, PhysRevA.21.588, PhysRevA.23.2397, PhysRevA.51.4010, PhysRevA.79.052504, Beterov_2009} and hence a reduction of its lifetime, which can be detrimental to applications in quantum information. Although several measurements of Rydberg state lifetimes have been performed \cite{PhysRevA.11.1504, doi:10.1063/1.435914, doi:10.1063/1.465011, doi:10.1063/1.467534, PhysRevA.65.031401, fengzhigang, PhysRevA.92.012517, PhysRevLett.37.1465, Nosbaum_1995, doi:10.1063/1.1396856, PhysRevA.77.052712, Branden_2009, FANG2001469, Du:21}, a thorough and quantitative investigation of the effects of blackbody radiation on high-lying Rydberg states (above $n\approx 60$), which are often used in such applications, has not been reported in the literature thus far. 

Here we present measurements of blackbody radiation (BBR)-induced transition rates from high-lying $nS$, $nP$ and $nD$ Rydberg states with principal quantum number $60 \leqslant n \leqslant 110$ of rubidium atoms in a magneto-optical trap (MOT). Such measurements require a method that allows one to distinguish between a target $nL$ state and nearby $n^\prime L^\prime $ states, which can be as close as a few $\mathrm{GHz}$ in frequency units. In this paper, we use a method we recently presented in \cite{PhysRevA.100.030501} which combines field ionization with state-selective depumping \cite{PhysRevA.96.043411} to measure the lifetimes of single Rydberg states and of state-ensembles (i.e., all the Rydberg levels that are populated by BBR), from which transition rates can be calculated. Furthermore, this method allow us also to observe the evolution of the population of the support states, i.e., all the populated Rydberg states except the target one.

We find that our experimental results deviate significantly from theoretical calculations \cite{PhysRevLett.42.835, PhysRevA.21.588, PhysRevA.23.2397, PhysRevA.51.4010, PhysRevA.79.052504, Beterov_2009} in well-defined ranges of $n$. We attribute these deviations to the spectral intensity distribution of the BBR within the apparatus \cite{baltes, PhysRevA.89.013847, PhysRevA.87.033801, PhysRevA.78.023806, PhysRevA.1.1170, kim, HAROCHE1985347}. Placing additional electrodes around the MOT leads to further changes in the observed transition rates, which indicates that under suitable conditions BBR-induced transitions could also be suppressed in our setup, thus increasing the lifetimes of the Rydberg states without the need for cooling down the apparatus \cite{Magnani_2020}. This is confirmed by simple model calculations that take into account the additional electrodes.

\section{\label{sec:level2} Experiment}
\begin{figure*}
\includegraphics[width=\textwidth]{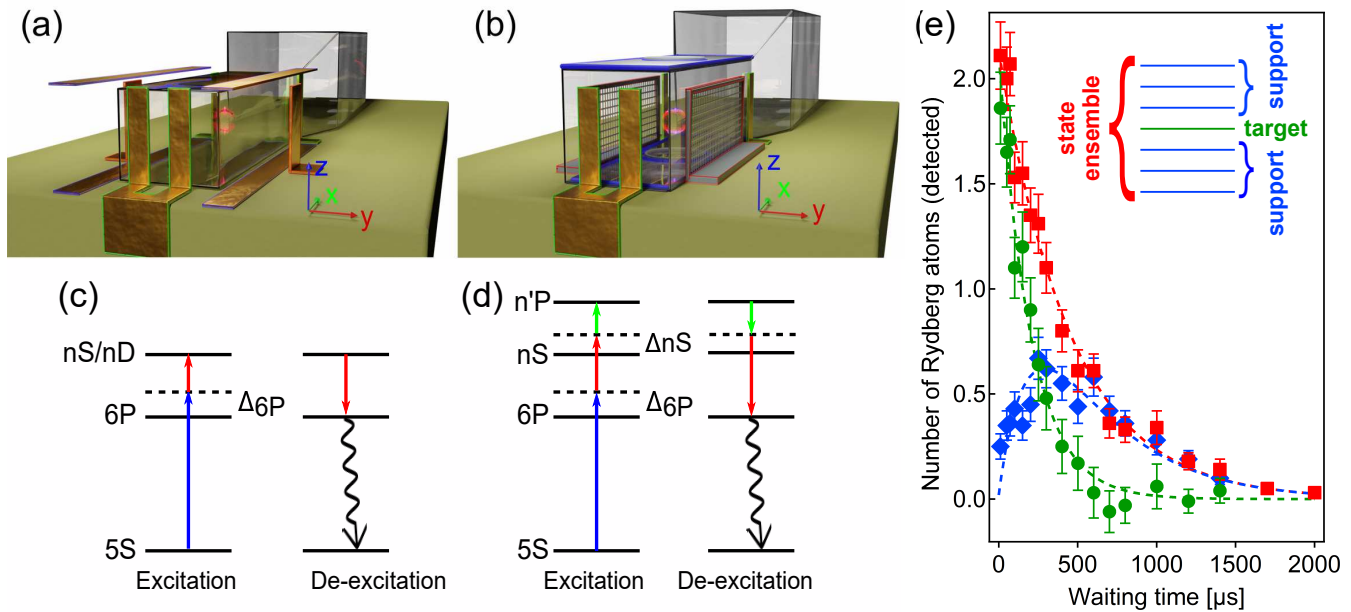}
\caption{\label{fig:0} Experimental setup and protocols for lifetime measurements. (a-b) shows 3D models of the quartz cell and electrodes used for the compensation of the stray electric field. In the first configuration (a), two pairs of brass electrodes are placed above and below the cell and extend along its length, while the frontal and lateral electrodes (which are also used for field ionization) are used for the compensation along the cell main axis. A pair of brass electrodes is placed on the sides of the cell to compensate the field along the y-axis. In a second configuration (b), two lateral wire mesh electrodes are placed along the sides, while in the vertical direction, aluminium electrodes with a circular hole in the centre covered with wire mesh are placed on the top and bottom of the cell. Each electrode pair is highlighted with the colour of the respective axis on which it generates a potential difference: green for the x-axis, red for the y-axis and blue for the z-axis. (c) Excitation and de-excitation scheme for $S$ and $D$ Rydberg states, in which the blue and red arrows represent the $420\,\mathrm{nm}$ and $1013\,\mathrm{nm}$ laser light, respectively, while the curly arrow indicates the fast radiative decay to the ground state. (d) Excitation and de-excitation for $P$ Rydberg states, for which an additional microwave field was used (green arrow). (e) A typical experimental measurement of the lifetime of a Rydberg state ($80S$): state ensemble (red squares), support (blue diamonds) and target state (green circles). The red and green dashed lines are single exponential fits to the state ensemble and target state data, respectively, while the blue dashed line is obtained from the difference of these two fits. The inset of the figure shows a schematic representation of the Rydberg levels involved in the dynamics.}
\end{figure*}

Our measurements of BBR-induced Rydberg-Rydberg transition rates are based on a combination of state-selective depumping \cite{PhysRevA.96.043411} and field ionization, as described in detail in \cite{PhysRevA.100.030501}. A schematic of the protocol is reported in Fig. \ref{fig:0}(c, d). Briefly, an $nL$ Rydberg target state is initially excited and after a variable waiting time $t$ either the state-ensemble population is measured by simple field ionization, or the support population is measured by first depumping the initially excited target state via the intermediate $6P_{3/2}$ state and then field ionizing the remaining Rydberg atoms. The target state, the state-ensemble and the support states are schematically represented in the inset of Fig. \ref{fig:0}(e). Field ionization is performed by applying a voltage difference between $4.5$ and $6.5\,\mathrm{kV}$ to two sets of electrodes located at opposite ends of the glass cell, which creates electric fields between $21\,\mathrm{V/cm}$ and $29\,\mathrm{V/cm}$ at the position of the MOT. The ions thus created are then detected by a channel electron multiplier (Channeltron). From the difference of the two populations (state-ensemble and support state), we finally calculate the target state population at time $t$ and the Rydberg-Rydberg transition rates (explained in detail below). Fig. \ref{fig:0}(e) shows a typical lifetime measurement for $80S$ Rydberg state.

In this work, we performed those measurements for $nS$, $nP$ and $nD$ states in a range of $n$ from $60$ to $110$. Rubidium atoms (${}^{87} Rb$) are initially collected in a standard MOT containing around $2\times 10^{5}$ atoms at $150\,\mathrm{\mu K}$ in a roughly spherical volume of about $300\,\mathrm{\mu m}$ in diameter. The MOT is formed inside a vacuum cell made of Vycor glass (detailed dimensions below). Fig. \ref{fig:0}(a) shows a schematic 3D model of the experimental setup. A thorough description of the apparatus can be found in \cite{Viteau_2010, Viteau:11}. Rydberg states are then excited through a combination of laser pulses at $420\,\mathrm{nm}$ and around $1013\,\mathrm{nm}$ (with laser beam diameters of $40$ and $90\,\mathrm{\mu m}$, respectively) and microwave pulses. 

For the $S$ and $D$ states, excitation is achieved by a two-photon process via the intermediate $6P_{3/2}$ state, from which the first photon at $420\,\mathrm{nm}$ is detuned by $36\,\mathrm{MHz}$. The two-photon Rabi frequency is approximately $\Omega= 2\pi \times 1\,\mathrm{MHz}$. To excite $n^\prime P$ states, (where $n^\prime $ is equal to $n$ or $n\pm 1$ with $n$ being the principal quantum number of the intermediate $S$ state), we use an additional microwave photon with a frequency between $1$ and $15\,\mathrm{GHz}$, which is delivered via a helical antenna mounted around $50\,\mathrm{cm}$ away from the glass cell in which the MOT is formed. Excitation of the $n^\prime P$ state is achieved through a three-photon process via the $nS$ state, where the $1013\,\mathrm{nm}$ laser is detuned by about $28\,\mathrm{MHz}$ from that state to avoid populating it, and finally to the $n^\prime P$ state with a microwave photon (the Rabi frequency of the last step is around $1-5\,\mathrm{MHz}$, measured via the Autler-Townes splitting of the $nS$ state when the microwave is in resonance with the $nS-n^\prime P$ transition). In order to minimize effects due to van der Waals interaction between the Rydberg atoms \cite{PhysRevA.93.030701, PhysRevLett.110.263201} (which could affect the depumping efficiency) or collective effects, the excitation parameters are adjusted such as to create only around 3 Rydberg atoms at $t=0$, with a mean spacing between them of around $100\,\mathrm{\mu m}$. 

For the depumping step, in the case of $S$ and $D$ states a $5\,\mathrm{\mu s}$-pulse of only the $1013\,\mathrm{nm}$ beam, now shifted into resonance with the $6P_{3/2}-nS$ transition using the acousto-optic modulator in the beam path, is used to de-excite the target state. The beam intensity is chosen such as to obtain a Rabi frequency of around $1\,\mathrm{MHz}$, which is smaller than the inverse lifetime of the $6P_{3/2}$ level (around $120\,\mathrm{ns}$) in order to avoid coherent dynamics, so that the net effect of the de-excitation and subsequent spontaneous decay is a depumping down to the ground state. The efficiency of this depumping procedure is typically larger than $95\%$.

In the case of $P$ states, de-excitation to the $6P_{3/2}$ level is achieved using a combined pulse of the microwave and $1013\,\mathrm{nm}$ beam similarly to the excitation protocol, but with the acousto-optic modulator frequency adjusted such as to tune the two-photon transition $6P_{3/2}-nP$ into resonance (see Fig. \ref{fig:0}(d)). To obtain reasonable depumping efficiencies between $45$ and $90\%$ (depending on the state and hence also the available microwave power at the respective frequencies) we use a $10\,\mathrm{\mu s}$ depumping pulse for the $P$ states. In separate experiments using $S$ states, we verified that depumping efficiencies as low as $40\%$ still allow us to extract the target state lifetime (see below) with reasonable precision, and also that a $10\,\mathrm{\mu s}$ pulse is still short enough so as not to lead to large systematic errors in the lifetime measurement. 

Considerable care has to be taken to compensate for stray electric fields inside the glass cell, which in our apparatus are typically of the order of $100-200\,\mathrm{mV/cm}$, with day-to-day variations of up to $50\,\mathrm{mV/cm}$, the exact origins of which are unknown (we have observed, for instance, that dust particles settling on the glass cell can lead to field variations of several tens of $\mathrm{mV/cm}$). The static stray fields are cancelled by applying an electric potential of up to $30\,\mathrm{V}$ to the various electrodes surrounding the glass cell (see Fig. \ref{fig:0}(b)) and monitoring the resulting Stark shift of an $nS$ Rydberg state (typically between $n=90$ and $n=105$) as well as the separation of that state from the adjacent $n-3$ manifold (for details see the Appendix). In this way, we are able to compensate the stray fields to within $15\,\mathrm{mV/cm}$, allowing us to work with principal quantum numbers up to 110, for which the Inglis-Teller limit is $33\,\mathrm{mV/cm}$ (when the stray fields approach or exceed the Inglis-Teller limit, the measured single-state lifetimes drop sharply and no longer reflect the actual lifetimes of the target states \cite{PhysRevA.100.030501}).

\section{\label{sec:level3} Results}
\begin{figure}
\includegraphics[width=0.5\textwidth]{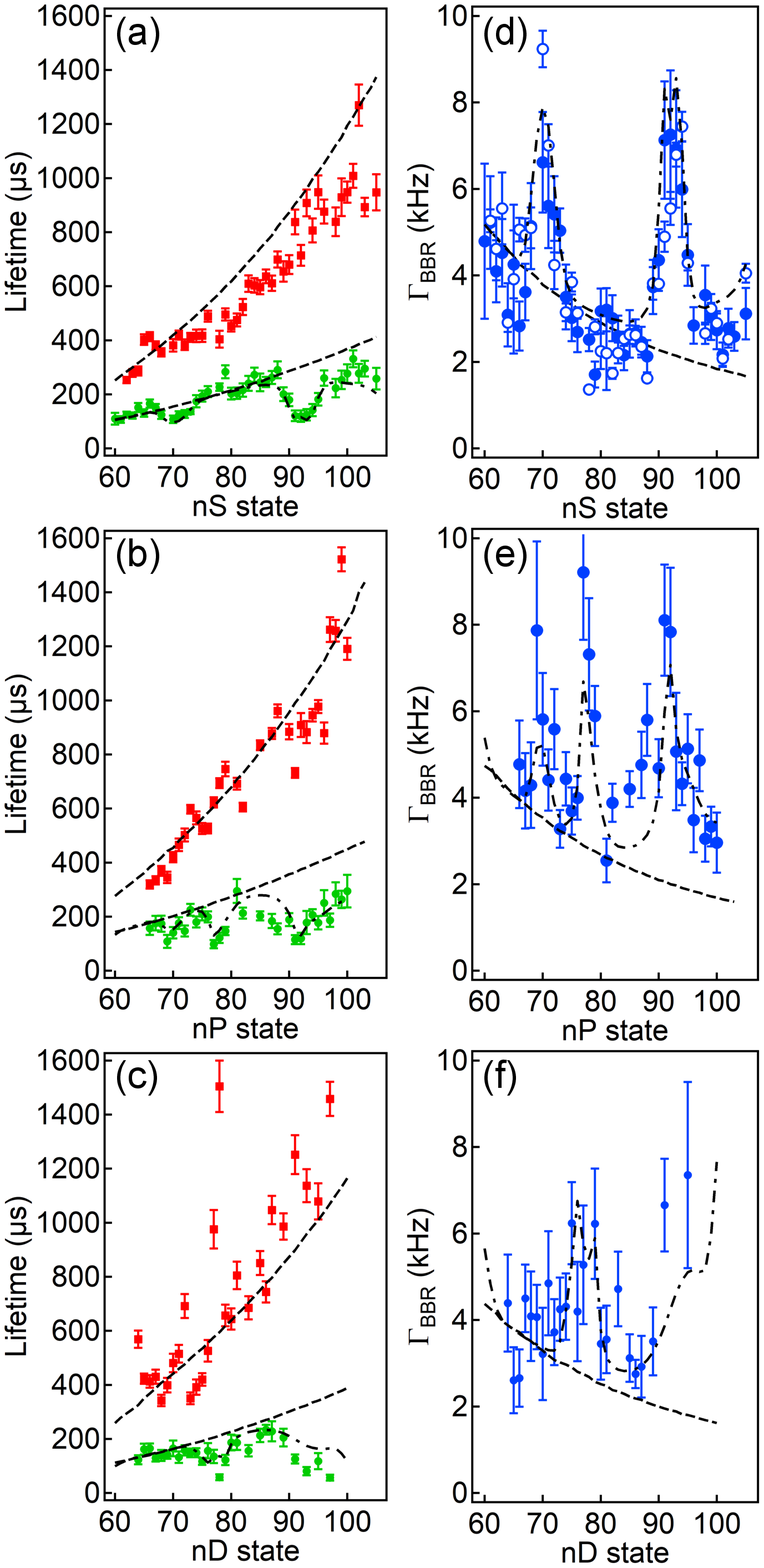}
\caption{\label{fig:1} Experimental results for state-ensemble and target-state lifetimes. (a-c) $S$, $P$ and $D$ state-ensemble (red squares) and target state lifetimes (green circles). Theoretical lifetimes (dashed black lines) and fits to the target state lifetimes using our modified BBR spectrum model (dash-dotted black line) are also shown. (d-f) Total BBR rates obtained from the experimental lifetimes for $S$, $P$ and $D$ states (blue circles), theoretical BBR rates (dashed black line) and BBR rates obtained from our model (dash-dotted black line). In (d) the BBR rates obtained by fitting the initial part of the support-state curve (open blue circles) are also shown.}
\end{figure}

Experimental results for state-ensemble and target-state lifetimes of $S$, $P$ and $D$ states are shown in Fig. \ref{fig:1}, along with numerical simulations based on lifetime calculations using the open-source software package ARC \cite{SIBALIC2017319, ROBERTSON2021107814}. Those lifetimes are extracted by fitting a single exponential decay curve (with offset fixed at $0$) to the experimental data. In the simulations, the time evolution of the target $nL$ level and its support states (with $n^\prime$ between $55$ and $120$ and  $L^\prime$ up to $4$, i.e., up to $G$ levels) is calculated by numerically integrating the coupled differential equations for those states. We note here that this leads to target state lifetimes that differ from the values usually quoted in the literature, as our simulations also take into account the re-population of the target state from support states. From the simulations, we evaluated this re-population rate to be around few percent of the total BBR transition rate between target and support states.

We report the experimental lifetimes relative to Rydberg states with the same principal $n$ and angular momentum $L$ quantum numbers but different values of the total angular momentum $J$ in the same data set, since the error bars on our measurements are of the same order of magnitude as the difference between those lifetimes (around a few percent). For the same reason, we report the theoretical values of the lifetimes relative to only one total angular momentum $J$, in particular $J=3/2$ for $P$ states and $J=5/2$ for $D$ states. 

As already observed in \cite{PhysRevA.100.030501}, we find good agreement between the experimental and theoretical state-ensemble lifetimes. For the $D$ states, the discrepancy between experiment and theory is larger, with more scatter in the data, which we attribute partly to the very large sensitivity to electric fields of the $F$ states, to which the $D$ states can couple. For the target state lifetimes, an agreement between the measured and theoretically predicted values (\cite{SIBALIC2017319, ROBERTSON2021107814}, which are based on the standard treatment of BBR, i.e., without cavity effects) is less good, with pronounced regions of shorter lifetimes in all three sets of states, especially at high principal quantum numbers.
   
To obtain a qualitative understanding of the observed deviations towards shorter target state lifetimes (which are the dominant effect; deviations towards longer lifetimes will be discussed below), we assume that there are regions in the microwave spectrum seen by the atoms for which the spectral intensity, and hence the corresponding Rydberg-Rydberg transition rate, is enhanced with respect to the value expected from a pure BBR microwave spectrum following Planck’s formula. Taking three such regions with centre frequencies $\nu_{i}$, Lorentzian widths $\Delta \nu_{i}$ and enhancement factor $\alpha_{i}$, we calculate the expected target state lifetimes of the $nS$ and $nP$ states for a given combination of the three parameters and then perform a simultaneous least-squares fit to the two data sets. The resulting target state lifetimes for $nS$, $nP$ and $nD$ states (for the latter we use the parameters obtained from the fit to the $S$ and $P$ states) are shown in Fig. \ref{fig:1}. For all three sets of states the calculated lifetimes agree well with the experimental data, indicating that the assumption of a limited number of well-defined frequency regions in which the atoms experience an enhanced spectral intensity of the microwave radiation is reasonable. We defer a discussion of the possible origin of those deviations to section \ref{sec:level4}.

Ultimately, we are interested in the BBR-induced transition rates $\Gamma_{BBR}$ between the target $nL$ Rydberg state and adjacent $n^\prime L^\prime$ states. Those rates can be extracted from the experimental data in two ways. In the first approach, we calculate the total decay rate $\Gamma_{tot} = 1/\tau_{target}$ of the target state from its measured lifetime and then obtain $\Gamma_{BBR}$ by subtracting its spontaneous decay rate $\Gamma_{spont} = 1/\tau_{spont}$, where the spontaneous decay lifetime $\tau_{spont}$ is calculated using ARC. Alternatively, we can extract $\Gamma_{BBR}$ from the experimental data without further input from theory by fitting a linear function to the first $100\,\mathrm{\mu s}$ of the support population curve and normalizing the resulting slope by the initial number of Rydberg excitations (we choose $100\,\mathrm{\mu s}$ as this is sufficiently shorter than the typical timescale for the involved transitions).

In Fig. \ref{fig:1} (d-f) we plot $\Gamma_{BBR}$ for $S$, $P$ and $D$ states obtained by the first approach. For the $S$ states we also plot the values obtained by the second approach; the two sets agree to within the experimental error. In principle, the second approach can also be used for $P$ and $D$ states. However, in the case of $P$ states, for which the depumping efficiencies are considerably lower than for the $S$ states, we find that the agreement between the two methods is less good, which is to be expected since the population values obtained by field ionizing after depumping are contaminated by residual population in the target state (which, on the other hand, does not matter in the first method, as the resulting target-state population curve is simply multiplied by the depumping efficiency). For $D$ states, on the other hand, the experimental data is generally noisier, which makes the second approach less viable. 

\section{\label{sec:level4} Discussion}
\begin{figure}
\includegraphics[width=0.5\textwidth]{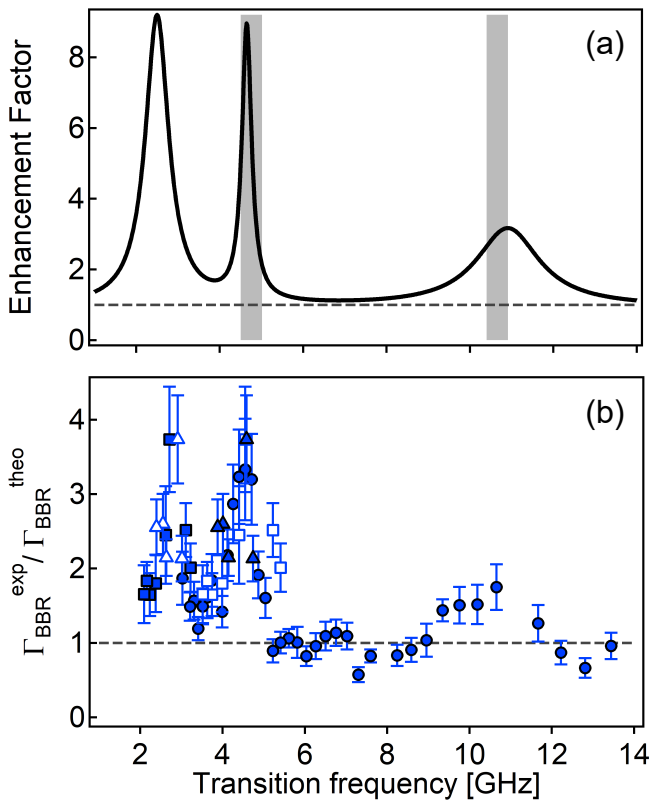}
\caption{\label{fig:2} (a) Enhancement factor with respect to the Planckian BBR spectrum as a function of frequency, consisting of three Lorentzian curves whose parameters are obtained by fitting the experimental target state lifetimes of the $S$ and $P$ states. The black dashed line indicates unity. Vertical grey bars centered at around $4.5\,\mathrm{GHz}$ and $10.4\,\mathrm{GHz}$ indicate the dominant features relative to the lifetime minima. (b) Ratios of experimental and theoretical BBR rates as a function of the frequency of the main BBR transitions: $nS_{1/2} \rightarrow nP_{3/2}$ (blue circles), $nP_{3/2} \rightarrow (n-1)D_{5/2}$ (blue squares), $nP_{1/2} \rightarrow nS_{1/2}$ (blue empty squares), $nP_{3/2}\rightarrow nS_{1/2}$ (blue triangles), and  $nP_{1/2} \rightarrow (n-1)D_{3/2}$ (blue empty triangles).}
\end{figure}

We now turn to the likely causes of the observed deviations from the theoretical predictions. In Fig. \ref{fig:2}(a) we plot the enhancement curve obtained from the fit described above and in Fig. \ref{fig:2}(b) the measured transition rates, normalized to the theoretical values (obtained using the software package ARC \cite{SIBALIC2017319, ROBERTSON2021107814}), from $nS$ and $nP$ states to the neighbouring support states. In the case of $nS$ states, we plot the measured rates as a function of the microwave frequency corresponding to the dominant transition (i.e., the one with the largest dipole moment) $nS-nP$. 

The dominant features around $n=71$ and $n=92$ correspond to transitions with frequencies around $10.4\,\mathrm{GHz}$ and $4.5\,\mathrm{GHz}$, respectively, reflected in peaks of the enhancement factor at those frequencies, as highlighted by the two vertical grey bars in Fig. \ref{fig:2} (a).
The peak at $2.5\,\mathrm{GHz}$, on the other hand, does not correspond to transitions from $S$ states but emerges when we plot the $P$-state measurements for $87 \leqslant n \leqslant 100$ as a function of the frequency corresponding to the transitions $nP \rightarrow (n-1)D$. We also plot $nP \rightarrow nS$ transitions, which have comparable rates to $nP  \rightarrow (n-1)D$ transitions, and confirm the enhancement around $4.5\,\mathrm{GHz}$ seen in the $nS \rightarrow nP$ rates.

A possible cause for such enhanced transition rates could, in principle, be the presence of external microwave signals reaching the atoms, for instance, due to WiFi, mobile phones or other microwave sources outside the apparatus. The observed increases in transition rates could be brought about by microwave signals with intensities as low as tens of $\,\mathrm{pW/cm^{2}}$. Those signals could either travel through the air or along conductors connected to electrodes or coils near the glass cell. In an attempt to rule out the former hypothesis, we shielded the apparatus by surrounding it from above with a cage made of aluminium foil of dimensions $2 \times 2 \times 1.5\,\mathrm{m}$ (shielding from below was provided by the optical table itself). 

We tested the shielding by placing a microwave antenna in a far corner of the laboratory, about 4 metres away from the apparatus and without a direct line of sight to the glass cell, to simulate external radiation arriving from different spatial directions. The microwave intensity reaching the atoms was measured via the Autler-Townes splitting of the $91S$ state with the microwave source tuned to the $91S-90P$ transition. By comparing that splitting with and without the shielding cage, we deduced a shielding factor of about $4$. We then measured the lifetime of the $91S$ state with and without the shielding, finding no difference within the experimental error. From this observation, we conclude that the observed enhanced transition rates in the frequency range corresponding to $nS$ states with $n$ around $90$ are not due to microwaves reaching the atoms directly from the outside, although we cannot rule out transmission through leads and other metallic structures. 

So far we have only discussed the enhancement of transition rates. However, particularly in the $S$ state data, we also observe regions of reduced transition rates. In Fig. \ref{fig:2}(b), these reductions of up to $30\%$ are evident around $7\,\mathrm{GHz}$ and $13\,\mathrm{GHz}$. This finding indicates an alternative explanation for the deviations: a modification of the spectral intensity of the BBR inside our apparatus \cite{baltes, PhysRevA.89.013847, PhysRevA.87.033801, PhysRevA.78.023806, PhysRevA.1.1170, kim, HAROCHE1985347}. Such a modification could be caused by a density of modes influenced by the presence of the cell walls and other structures such as supports and field coils close to the cell. The cell itself has a cross-section of $2.4\,\mathrm{cm} \times 3.0\,\mathrm{cm}$ (outside) and $1.8\,\mathrm{cm} \times 2.4\,\mathrm{cm}$ (inside), and the nearest surrounding structures have a spacing between them (roughly symmetrically about the centre of the cell, which is where the MOT is formed) of around $3-5\,\mathrm{cm}$. Assuming that we are seeing an enhancement of the transition rates due to an enhanced density of modes that coincides with the conditions for standing waves with a wavelength around $3.6-10\,\mathrm{cm}$ (i.e., twice the dimensions quoted above for the fundamental mode), we would expect to find the deviations for frequencies between $3\,\mathrm{GHz}$ and $8\,\mathrm{GHz}$ and higher harmonics thereof, which agrees quite well with our observations (see Fig. 3(a) and (b)). 

Together with the observed regions of suppressed BBR rates, this strongly indicates a modified BBR spectrum. 
To further test this hypothesis, we add electrodes around the glass cell in order to modify the density of modes (see Fig.\ref{fig:0}(b)). In the horizontal direction, two lateral wire mesh electrodes of dimensions $1.8\,\mathrm{cm} \times 6.0\,\mathrm{cm}$ are placed at a variable distance to the left and right of the cell, allowing the MOT beams to pass through with less than $20\%$ attenuation whilst almost completely ($>90\%$) reflecting microwaves in the frequency range of interest (around $5\,\mathrm{GHz}$). In the vertical direction, aluminium electrodes of dimensions $3.0\,\mathrm{cm} \times 9.0\,\mathrm{cm}$ are placed on the top and bottom of the cell. At the centre of the electrodes, a $2.1\,\mathrm{cm}$ hole covered by wire mesh allows the vertical MOT beams to pass through. Additionally, the supporting structures laterally above and below the cell, which have a vertical separation of $2.1\,\mathrm{cm}$, are covered with aluminium foil.

\begin{figure}
\includegraphics[width=0.45\textwidth]{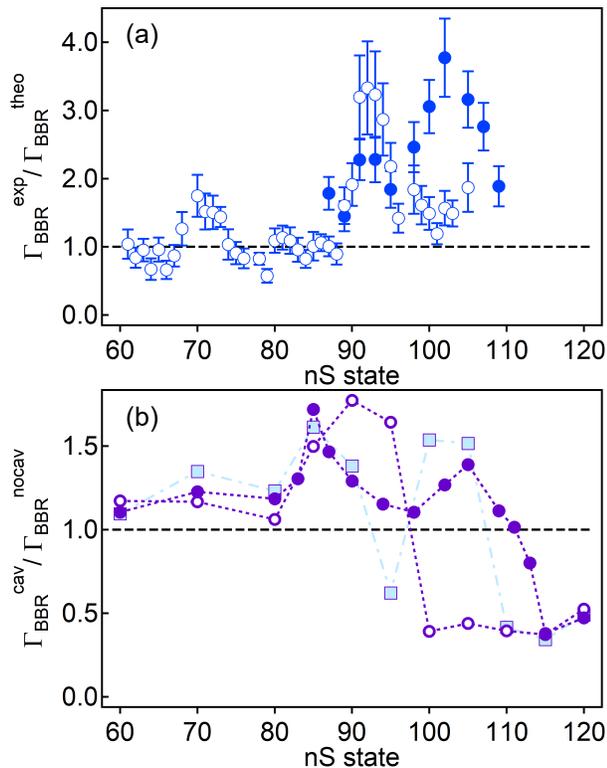}
\caption{\label{fig:3} (a) Normalized experimental rates $\Gamma_{BBR}$ obtained from the measured target state lifetimes for $nS$ states without (empty blue circles) and with additional electrodes (solid blue circles) with a $42\,\mathrm{mm}$ separation between the additional lateral electrodes. (b) Numerical simulation of the normalized BBR rates for $nS$ Rydberg states without any cavity (horizontal dashed line) and with additional electrodes with separations $30\,\mathrm{mm} \times 26\,\mathrm{mm} \times 300\,\mathrm{mm}$ (empty purple circles),  $42\,\mathrm{mm} \times 26\,\mathrm{mm} \times 300\,\mathrm{mm}$ (solid purple circles), and  with a cavity of the same dimension as the latter but with a higher Q-value (light blue squares).}
\end{figure}
Fig. \ref{fig:3}(a) shows the measured BBR rates with the addition of those electrodes, with a $4.2\,\mathrm{cm}$ separation between the lateral electrodes (symmetrically around the centre of the cell). For comparison, the transition rates measured without the additional electrodes are also shown. The two curves are clearly different, indicating that the addition of electrodes changed the electromagnetic mode structure inside the cell. The presence of the glass cell, which we assume to have a refractive index around $2.5$ at frequencies of a few $\mathrm{GHz}$ \cite{Birch_1975}, makes it difficult to predict the exact mode distribution due to the horizontal and vertical electrodes (based on a refractive index of $2.5$, the expected reflectivity is $18\%$). For simplicity, we calculate an approximate mode distribution by assuming an average refractive index taking into account the thickness of the two glass walls in each direction. For the direction along the main axis of the cell, we assume a characteristic length scale of $30\,\mathrm{cm}$ (defined by the front cell wall and the back of the vacuum chamber to which the cell is attached).
The resulting allowed modes in the cavity are counted and summed over bins of width $500\,\mathrm{MHz}$ to account for the low expected Q factor, which we take to be $10$. This mode density distribution is then used in the numerical simulation described above to calculate the ensemble, support and target populations as a function of time. BBR-induced transition rates from the target state are calculated from fits to the target state populations (analogously to the first approach used for the experimental data, see above). The resulting transition rates are normalized to the case of a standard Planckian mode density distribution. 

Fig. \ref{fig:3}(b) shows the normalized transition rates calculated in this way for two configurations of the lateral electrodes. In both configurations, local maxima of the normalized rates are visible, whose positions agree roughly with those of the experimentally observed maxima. Above $n=100$ and $n=110$, respectively, the normalized rate drops below $1$, indicating the low-frequency cut-off of the cavity. Experimentally, we do not observe such a drop up to $n=110$. Given the uncertainties in our measurements of the separation between the electrodes ($\pm 1\,\mathrm{mm}$) and in the estimate of the resonant frequencies based on the simplistic model of an effective refractive index to take into account the presence of the glass cell, it is conceivable that the value of $n$ corresponding to the actual cut-off is higher than the model suggests. Currently, we are unable to explore states with $n>110$ because of the minimum residual electric fields we can achieve in a stable manner using field compensation. In principle, however, it should be possible to observe drastically reduced BBR-induced transition rates in a setup like ours \cite{Magnani_2020}.
\section{\label{sec:level5} Conclusion}
In summary, in this work we investigated the transition rates between high-lying Rydberg states of rubidium atoms due to BBR over a wide range of principal quantum numbers and for different angular momenta. The deviations of the measured transition rates from theoretical predictions suggest that the glass cell and other structures in our apparatus that reflect microwaves act as a low-Q microwave cavity and hence modify the density of electromagnetic modes, leading to a departure from the Planckian BBR spectrum. To test this hypothesis, we placed additional electrodes around the vacuum cell in order to further modify the mode density. We observed a clear variation in the transition rates which agrees well with a simple model for the spectral density of modes with and without the additional electrodes. That model also suggests that above a critical principal quantum number the transition rates should be strongly suppressed because the relevant microwave transition frequencies are below the cut-off frequency of the cavity. While we were unable to reach that regime in the present work, it should be possible in general to control and suppress the BBR-induced transitions rates in Rydberg experiments, and hence increase Rydberg state lifetimes, using an additional microwave cavity. Our findings are relevant to a number of applications of Rydberg atoms, such as in Rydberg quantum thermometry \cite{Norrgard_2021, PhysRevLett.107.093003}, where precise knowledge of the BBR-induced transition rates is relevant. Also, the possibility of reducing BBR-induced transition rates suggests the potential for enhancing the performance of Rydberg atoms in quantum computation, quantum simulation and quantum metrology \cite{ZollerNature2010, PhysRevX.8.011032, BrowaeysNature2020, RevModPhys.82.2313, Adams_2019}.
\section{Appendix}
\appendix
\section{Compensation technique}
The presence of a stray electric field in the region where the MOT is formed greatly affects the measurement of the lifetime of high-lying Rydberg states, as we showed in our previous work \cite{PhysRevA.100.030501}. In fact, due to their high polarizability, even electric fields as small as few hundreds of $\mathrm{mV/cm}$ are sufficient to mix the target state with the neighbouring manifold of states with $l\geqslant 3$. The magnitude of the stray electric field in our setup is about $200\,\mathrm{mV/cm}$, with daily variations of about $50\,\mathrm{mV/cm}$.  

\begin{figure}
\centering
   \vspace{20pt}
	\includegraphics[width=0.45\textwidth]{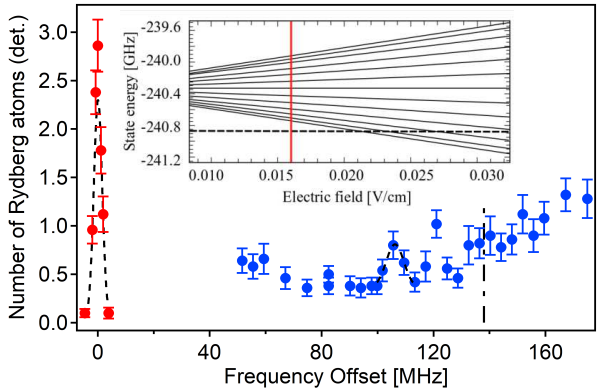}
	\caption{An excitation frequency scan highlights the resonance of the target Rydberg state $120S_{1/2}$ state (red circles) and the multiple resonances relative to the manifold of states with $l\geqslant 3$ (blue circles). Dashed black lines are fits to the target state and lowest manifold state resonance. The dash-dotted black line indicates the position of the second resonance of the manifold as calculated from the Stark map, which cannot be resolved from the reported data. By comparing the frequency offset, about $105\,\mathrm{MHz}$, between these two resonances with the calculated Stark map (inset), we obtain the value of the residual field (red line on the inset), which in this case is $16\pm 1\,\mathrm{mV/cm}$. Note that for exciting the manifold states we increased the intensity of the $1013\,\mathrm{nm}$ laser by a factor of around $80$ with respect to the excitation of the $120S$ state; the increase in the number of detected Rydberg atoms below a frequency offset of around $80\,\mathrm{MHz}$ is due to off-resonant excitation of the $120S$ state.}
	\label{model_compensation2}
\end{figure}
In our experiments, we compensate the background electric field during the excitation-detection cycle (lasting around $3\,\mathrm{ms}$ in total) through the application of potentials to a set of electrodes placed around the quartz cell. In the first configuration (see Fig. \ref{fig:0}(a)), two pairs of brass electrodes were placed above and below the cell and extended along its length to create an electric field along the vertical direction (z-axis), while the frontal and lateral electrodes, also used for the application of the ionization field, were used for the compensation along the main axis of the cell (x-axis). A pair of brass electrodes is placed on the sides of the cell to compensate the field along the y-axis. These elements were subsequently replaced with the mesh electrodes used to study the change of the mode density inside the cell, as described in Section \ref{sec:level4} (see Fig. \ref{fig:0}(b)). This configuration allows controlling the electric field in the three orthogonal directions.

The compensation electric field is determined by changing the voltages applied to the electrodes and iteratively minimizing the Stark shift. The compensation procedure is typically performed with a high-lying state such as $120S_{1/2}$ to maximize the sensitivity. The magnitude of the residual electric field present after the compensation procedure can be obtained by measuring the frequency difference between the target state resonance and the first resonance of the manifold of states with $l\geqslant 3$, and comparing the resulting value with the Stark map calculated using the ARC software package (see inset of Fig. \ref{model_compensation2}). For the measurement of the manifold states, the power of the excitation lasers is increased in order to obtain a sufficiently large signal. 
This method allows us to reduce the magnitude of the stray electric field down to $15\,\mathrm{mV/cm}$, permitting the measurement of the lifetimes of Rydberg states with a principal quantum number up to $n=110$ (for the lifetime experiments we use that upper limit to make sure that a possible increase in the background field after the compensation procedure does not lead to errors in the measurements).

\bibliography{RydBBRRateDraft}

\providecommand{\noopsort}[1]{}\providecommand{\singleletter}[1]{#1}%
\begin{thebibliography}{54}%
\makeatletter
\providecommand \@ifxundefined [1]{%
 \@ifx{#1\undefined}
}%
\providecommand \@ifnum [1]{%
 \ifnum #1\expandafter \@firstoftwo
 \else \expandafter \@secondoftwo
 \fi
}%
\providecommand \@ifx [1]{%
 \ifx #1\expandafter \@firstoftwo
 \else \expandafter \@secondoftwo
 \fi
}%
\providecommand \natexlab [1]{#1}%
\providecommand \enquote  [1]{``#1''}%
\providecommand \bibnamefont  [1]{#1}%
\providecommand \bibfnamefont [1]{#1}%
\providecommand \citenamefont [1]{#1}%
\providecommand \href@noop [0]{\@secondoftwo}%
\providecommand \href [0]{\begingroup \@sanitize@url \@href}%
\providecommand \@href[1]{\@@startlink{#1}\@@href}%
\providecommand \@@href[1]{\endgroup#1\@@endlink}%
\providecommand \@sanitize@url [0]{\catcode `\\12\catcode `\$12\catcode
  `\&12\catcode `\#12\catcode `\^12\catcode `\_12\catcode `\%12\relax}%
\providecommand \@@startlink[1]{}%
\providecommand \@@endlink[0]{}%
\providecommand \url  [0]{\begingroup\@sanitize@url \@url }%
\providecommand \@url [1]{\endgroup\@href {#1}{\urlprefix }}%
\providecommand \urlprefix  [0]{URL }%
\providecommand \Eprint [0]{\href }%
\providecommand \doibase [0]{https://doi.org/}%
\providecommand \selectlanguage [0]{\@gobble}%
\providecommand \bibinfo  [0]{\@secondoftwo}%
\providecommand \bibfield  [0]{\@secondoftwo}%
\providecommand \translation [1]{[#1]}%
\providecommand \BibitemOpen [0]{}%
\providecommand \bibitemStop [0]{}%
\providecommand \bibitemNoStop [0]{.\EOS\space}%
\providecommand \EOS [0]{\spacefactor3000\relax}%
\providecommand \BibitemShut  [1]{\csname bibitem#1\endcsname}%
\let\auto@bib@innerbib\@empty
\bibitem [{\citenamefont {Archimi}\ \emph {et~al.}(2019)\citenamefont
  {Archimi}, \citenamefont {Simonelli}, \citenamefont {Di~Virgilio},
  \citenamefont {Greco}, \citenamefont {Ceccanti}, \citenamefont {Arimondo},
  \citenamefont {Ciampini}, \citenamefont {Ryabtsev}, \citenamefont {Beterov},\
  and\ \citenamefont {Morsch}}]{PhysRevA.100.030501}%
  \BibitemOpen
  \bibfield  {author} {\bibinfo {author} {\bibfnamefont {M.}~\bibnamefont
  {Archimi}}, \bibinfo {author} {\bibfnamefont {C.}~\bibnamefont {Simonelli}},
  \bibinfo {author} {\bibfnamefont {L.}~\bibnamefont {Di~Virgilio}}, \bibinfo
  {author} {\bibfnamefont {A.}~\bibnamefont {Greco}}, \bibinfo {author}
  {\bibfnamefont {M.}~\bibnamefont {Ceccanti}}, \bibinfo {author}
  {\bibfnamefont {E.}~\bibnamefont {Arimondo}}, \bibinfo {author}
  {\bibfnamefont {D.}~\bibnamefont {Ciampini}}, \bibinfo {author}
  {\bibfnamefont {I.~I.}\ \bibnamefont {Ryabtsev}}, \bibinfo {author}
  {\bibfnamefont {I.~I.}\ \bibnamefont {Beterov}},\ and\ \bibinfo {author}
  {\bibfnamefont {O.}~\bibnamefont {Morsch}},\ }\bibfield  {title} {\bibinfo
  {title} {Measurements of single-state and state-ensemble lifetimes of
  high-lying {Rb} {Rydberg} levels},\ }\href
  {https://doi.org/10.1103/PhysRevA.100.030501} {\bibfield  {journal} {\bibinfo
   {journal} {Phys. Rev. A (R)}\ }\textbf {\bibinfo {volume} {100}},\ \bibinfo
  {pages} {030501} (\bibinfo {year} {2019})}\BibitemShut {NoStop}%
\bibitem [{\citenamefont {Simonelli}\ \emph {et~al.}(2017)\citenamefont
  {Simonelli}, \citenamefont {Archimi}, \citenamefont {Asteria}, \citenamefont
  {Capecchi}, \citenamefont {Masella}, \citenamefont {Arimondo}, \citenamefont
  {Ciampini},\ and\ \citenamefont {Morsch}}]{PhysRevA.96.043411}%
  \BibitemOpen
  \bibfield  {author} {\bibinfo {author} {\bibfnamefont {C.}~\bibnamefont
  {Simonelli}}, \bibinfo {author} {\bibfnamefont {M.}~\bibnamefont {Archimi}},
  \bibinfo {author} {\bibfnamefont {L.}~\bibnamefont {Asteria}}, \bibinfo
  {author} {\bibfnamefont {D.}~\bibnamefont {Capecchi}}, \bibinfo {author}
  {\bibfnamefont {G.}~\bibnamefont {Masella}}, \bibinfo {author} {\bibfnamefont
  {E.}~\bibnamefont {Arimondo}}, \bibinfo {author} {\bibfnamefont
  {D.}~\bibnamefont {Ciampini}},\ and\ \bibinfo {author} {\bibfnamefont
  {O.}~\bibnamefont {Morsch}},\ }\bibfield  {title} {\bibinfo {title}
  {Deexcitation spectroscopy of strongly interacting {Rydberg} gases},\ }\href
  {https://doi.org/10.1103/PhysRevA.96.043411} {\bibfield  {journal} {\bibinfo
  {journal} {Phys. Rev. A}\ }\textbf {\bibinfo {volume} {96}},\ \bibinfo
  {pages} {043411} (\bibinfo {year} {2017})}\BibitemShut {NoStop}%
\bibitem [{\citenamefont {Sedlacek}\ \emph {et~al.}(2012)\citenamefont
  {Sedlacek}, \citenamefont {Schwettmann}, \citenamefont {Kübler},
  \citenamefont {Löw}, \citenamefont {Pfau},\ and\ \citenamefont
  {Shaffer}}]{PfauNature2012}%
  \BibitemOpen
  \bibfield  {author} {\bibinfo {author} {\bibfnamefont {J.~A.}\ \bibnamefont
  {Sedlacek}}, \bibinfo {author} {\bibfnamefont {A.}~\bibnamefont
  {Schwettmann}}, \bibinfo {author} {\bibfnamefont {H.}~\bibnamefont
  {Kübler}}, \bibinfo {author} {\bibfnamefont {R.}~\bibnamefont {Löw}},
  \bibinfo {author} {\bibfnamefont {T.}~\bibnamefont {Pfau}},\ and\ \bibinfo
  {author} {\bibfnamefont {J.~P.}\ \bibnamefont {Shaffer}},\ }\bibfield
  {title} {\bibinfo {title} {Microwave electrometry with {Rydberg} atoms in a
  vapour cell using bright atomic resonances},\ }\href
  {https://doi.org/10.1038/nphys2423} {\bibfield  {journal} {\bibinfo
  {journal} {Nature Phys.}\ }\textbf {\bibinfo {volume} {8}},\ \bibinfo {pages}
  {819} (\bibinfo {year} {2012})}\BibitemShut {NoStop}%
\bibitem [{\citenamefont {Osterwalder}\ and\ \citenamefont
  {Merkt}(1999)}]{PhysRevLett.82.1831}%
  \BibitemOpen
  \bibfield  {author} {\bibinfo {author} {\bibfnamefont {A.}~\bibnamefont
  {Osterwalder}}\ and\ \bibinfo {author} {\bibfnamefont {F.}~\bibnamefont
  {Merkt}},\ }\bibfield  {title} {\bibinfo {title} {Using high {Rydberg} states
  as electric field sensors},\ }\href
  {https://doi.org/10.1103/PhysRevLett.82.1831} {\bibfield  {journal} {\bibinfo
   {journal} {Phys. Rev. Lett.}\ }\textbf {\bibinfo {volume} {82}},\ \bibinfo
  {pages} {1831} (\bibinfo {year} {1999})}\BibitemShut {NoStop}%
\bibitem [{\citenamefont {Neukammer}\ \emph {et~al.}(1987)\citenamefont
  {Neukammer}, \citenamefont {Rinneberg}, \citenamefont {Vietzke},
  \citenamefont {K\"onig}, \citenamefont {Hieronymus}, \citenamefont {Kohl},
  \citenamefont {Grabka},\ and\ \citenamefont {Wunner}}]{PhysRevLett.59.2947}%
  \BibitemOpen
  \bibfield  {author} {\bibinfo {author} {\bibfnamefont {J.}~\bibnamefont
  {Neukammer}}, \bibinfo {author} {\bibfnamefont {H.}~\bibnamefont
  {Rinneberg}}, \bibinfo {author} {\bibfnamefont {K.}~\bibnamefont {Vietzke}},
  \bibinfo {author} {\bibfnamefont {A.}~\bibnamefont {K\"onig}}, \bibinfo
  {author} {\bibfnamefont {H.}~\bibnamefont {Hieronymus}}, \bibinfo {author}
  {\bibfnamefont {M.}~\bibnamefont {Kohl}}, \bibinfo {author} {\bibfnamefont
  {H.~J.}\ \bibnamefont {Grabka}},\ and\ \bibinfo {author} {\bibfnamefont
  {G.}~\bibnamefont {Wunner}},\ }\bibfield  {title} {\bibinfo {title}
  {Spectroscopy of {Rydberg} atoms at $n\ensuremath{\approx}500$: Observation
  of quasi-{L}andau resonances in low magnetic fields},\ }\href
  {https://doi.org/10.1103/PhysRevLett.59.2947} {\bibfield  {journal} {\bibinfo
   {journal} {Phys. Rev. Lett.}\ }\textbf {\bibinfo {volume} {59}},\ \bibinfo
  {pages} {2947} (\bibinfo {year} {1987})}\BibitemShut {NoStop}%
\bibitem [{\citenamefont {Fan}\ \emph {et~al.}(2015)\citenamefont {Fan},
  \citenamefont {Kumar}, \citenamefont {Sedlacek}, \citenamefont {Kübler},
  \citenamefont {Karimkashi},\ and\ \citenamefont {Shaffer}}]{Fan_2015}%
  \BibitemOpen
  \bibfield  {author} {\bibinfo {author} {\bibfnamefont {H.}~\bibnamefont
  {Fan}}, \bibinfo {author} {\bibfnamefont {S.}~\bibnamefont {Kumar}}, \bibinfo
  {author} {\bibfnamefont {J.}~\bibnamefont {Sedlacek}}, \bibinfo {author}
  {\bibfnamefont {H.}~\bibnamefont {Kübler}}, \bibinfo {author} {\bibfnamefont
  {S.}~\bibnamefont {Karimkashi}},\ and\ \bibinfo {author} {\bibfnamefont
  {J.~P.}\ \bibnamefont {Shaffer}},\ }\bibfield  {title} {\bibinfo {title}
  {Atom based {RF} electric field sensing},\ }\href
  {https://doi.org/10.1088/0953-4075/48/20/202001} {\bibfield  {journal}
  {\bibinfo  {journal} {J. Phys. B - At. Mol. Opt.}\ }\textbf {\bibinfo
  {volume} {48}},\ \bibinfo {pages} {202001} (\bibinfo {year}
  {2015})}\BibitemShut {NoStop}%
\bibitem [{\citenamefont {Larrouy}\ \emph {et~al.}(2019)\citenamefont
  {Larrouy}, \citenamefont {Dietsche}, \citenamefont {Richaud}, \citenamefont
  {Raimond}, \citenamefont {Brune},\ and\ \citenamefont
  {Gleyzes}}]{Larrouy:19}%
  \BibitemOpen
  \bibfield  {author} {\bibinfo {author} {\bibfnamefont {A.}~\bibnamefont
  {Larrouy}}, \bibinfo {author} {\bibfnamefont {E.~K.}\ \bibnamefont
  {Dietsche}}, \bibinfo {author} {\bibfnamefont {R.}~\bibnamefont {Richaud}},
  \bibinfo {author} {\bibfnamefont {J.~M.}\ \bibnamefont {Raimond}}, \bibinfo
  {author} {\bibfnamefont {M.}~\bibnamefont {Brune}},\ and\ \bibinfo {author}
  {\bibfnamefont {S.}~\bibnamefont {Gleyzes}},\ }\bibfield  {title} {\bibinfo
  {title} {Quantum sensing using {Rydberg} atoms},\ }in\ \href
  {https://doi.org/10.1364/QIM.2019.S3A.5} {\emph {\bibinfo {booktitle} {QIM V:
  Quantum Technologies}}}\ (\bibinfo  {publisher} {Optical Society of
  America},\ \bibinfo {year} {2019})\ p.\ \bibinfo {pages} {S3A.5}\BibitemShut
  {NoStop}%
\bibitem [{\citenamefont {{Holloway}}\ \emph {et~al.}(2014)\citenamefont
  {{Holloway}}, \citenamefont {{Gordon}}, \citenamefont {{Jefferts}},
  \citenamefont {{Schwarzkopf}}, \citenamefont {{Anderson}}, \citenamefont
  {{Miller}}, \citenamefont {{Thaicharoen}},\ and\ \citenamefont
  {{Raithel}}}]{6910267}%
  \BibitemOpen
  \bibfield  {author} {\bibinfo {author} {\bibfnamefont {C.~L.}\ \bibnamefont
  {{Holloway}}}, \bibinfo {author} {\bibfnamefont {J.~A.}\ \bibnamefont
  {{Gordon}}}, \bibinfo {author} {\bibfnamefont {S.}~\bibnamefont
  {{Jefferts}}}, \bibinfo {author} {\bibfnamefont {A.}~\bibnamefont
  {{Schwarzkopf}}}, \bibinfo {author} {\bibfnamefont {D.~A.}\ \bibnamefont
  {{Anderson}}}, \bibinfo {author} {\bibfnamefont {S.~A.}\ \bibnamefont
  {{Miller}}}, \bibinfo {author} {\bibfnamefont {N.}~\bibnamefont
  {{Thaicharoen}}},\ and\ \bibinfo {author} {\bibfnamefont {G.}~\bibnamefont
  {{Raithel}}},\ }\bibfield  {title} {\bibinfo {title} {Broadband {Rydberg}
  atom-based electric-field probe for si-traceable, self-calibrated
  measurements},\ }\href {https://doi.org/10.1109/TAP.2014.2360208} {\bibfield
  {journal} {\bibinfo  {journal} {IEEE Trans. Antennas Propag.}\ }\textbf
  {\bibinfo {volume} {62}},\ \bibinfo {pages} {6169} (\bibinfo {year}
  {2014})}\BibitemShut {NoStop}%
\bibitem [{\citenamefont {Saffman}\ \emph {et~al.}(2010)\citenamefont
  {Saffman}, \citenamefont {Walker},\ and\ \citenamefont
  {M\o{}lmer}}]{RevModPhys.82.2313}%
  \BibitemOpen
  \bibfield  {author} {\bibinfo {author} {\bibfnamefont {M.}~\bibnamefont
  {Saffman}}, \bibinfo {author} {\bibfnamefont {T.~G.}\ \bibnamefont
  {Walker}},\ and\ \bibinfo {author} {\bibfnamefont {K.}~\bibnamefont
  {M\o{}lmer}},\ }\bibfield  {title} {\bibinfo {title} {Quantum information
  with {Rydberg} atoms},\ }\href {https://doi.org/10.1103/RevModPhys.82.2313}
  {\bibfield  {journal} {\bibinfo  {journal} {Rev. Mod. Phys.}\ }\textbf
  {\bibinfo {volume} {82}},\ \bibinfo {pages} {2313} (\bibinfo {year}
  {2010})}\BibitemShut {NoStop}%
\bibitem [{\citenamefont {Weimer}\ \emph {et~al.}(2010)\citenamefont {Weimer},
  \citenamefont {Müller}, \citenamefont {Lesanovsky}, \citenamefont {Zoller},\
  and\ \citenamefont {Büchler}}]{ZollerNature2010}%
  \BibitemOpen
  \bibfield  {author} {\bibinfo {author} {\bibfnamefont {H.}~\bibnamefont
  {Weimer}}, \bibinfo {author} {\bibfnamefont {M.}~\bibnamefont {Müller}},
  \bibinfo {author} {\bibfnamefont {I.}~\bibnamefont {Lesanovsky}}, \bibinfo
  {author} {\bibfnamefont {P.}~\bibnamefont {Zoller}},\ and\ \bibinfo {author}
  {\bibfnamefont {H.~P.}\ \bibnamefont {Büchler}},\ }\bibfield  {title}
  {\bibinfo {title} {A {Rydberg} quantum simulator},\ }\href
  {https://doi.org/10.1038/nphys1614} {\bibfield  {journal} {\bibinfo
  {journal} {Nature Phys.}\ }\textbf {\bibinfo {volume} {6}},\ \bibinfo {pages}
  {382–} (\bibinfo {year} {2010})}\BibitemShut {NoStop}%
\bibitem [{\citenamefont {Nguyen}\ \emph {et~al.}(2018)\citenamefont {Nguyen},
  \citenamefont {Raimond}, \citenamefont {Sayrin}, \citenamefont {Corti\~nas},
  \citenamefont {Cantat-Moltrecht}, \citenamefont {Assemat}, \citenamefont
  {Dotsenko}, \citenamefont {Gleyzes}, \citenamefont {Haroche}, \citenamefont
  {Roux}, \citenamefont {Jolicoeur},\ and\ \citenamefont
  {Brune}}]{PhysRevX.8.011032}%
  \BibitemOpen
  \bibfield  {author} {\bibinfo {author} {\bibfnamefont {T.~L.}\ \bibnamefont
  {Nguyen}}, \bibinfo {author} {\bibfnamefont {J.~M.}\ \bibnamefont {Raimond}},
  \bibinfo {author} {\bibfnamefont {C.}~\bibnamefont {Sayrin}}, \bibinfo
  {author} {\bibfnamefont {R.}~\bibnamefont {Corti\~nas}}, \bibinfo {author}
  {\bibfnamefont {T.}~\bibnamefont {Cantat-Moltrecht}}, \bibinfo {author}
  {\bibfnamefont {F.}~\bibnamefont {Assemat}}, \bibinfo {author} {\bibfnamefont
  {I.}~\bibnamefont {Dotsenko}}, \bibinfo {author} {\bibfnamefont
  {S.}~\bibnamefont {Gleyzes}}, \bibinfo {author} {\bibfnamefont
  {S.}~\bibnamefont {Haroche}}, \bibinfo {author} {\bibfnamefont
  {G.}~\bibnamefont {Roux}}, \bibinfo {author} {\bibfnamefont {T.}~\bibnamefont
  {Jolicoeur}},\ and\ \bibinfo {author} {\bibfnamefont {M.}~\bibnamefont
  {Brune}},\ }\bibfield  {title} {\bibinfo {title} {Towards quantum simulation
  with circular {Rydberg} atoms},\ }\href
  {https://doi.org/10.1103/PhysRevX.8.011032} {\bibfield  {journal} {\bibinfo
  {journal} {Phys. Rev. X}\ }\textbf {\bibinfo {volume} {8}},\ \bibinfo {pages}
  {011032} (\bibinfo {year} {2018})}\BibitemShut {NoStop}%
\bibitem [{\citenamefont {Browaeys}\ and\ \citenamefont
  {Lahaye}(2020)}]{BrowaeysNature2020}%
  \BibitemOpen
  \bibfield  {author} {\bibinfo {author} {\bibfnamefont {A.}~\bibnamefont
  {Browaeys}}\ and\ \bibinfo {author} {\bibfnamefont {T.}~\bibnamefont
  {Lahaye}},\ }\bibfield  {title} {\bibinfo {title} {Many-body physics with
  individually controlled {Rydberg} atoms},\ }\href
  {https://doi.org/10.1038/s41567-019-0733-z} {\bibfield  {journal} {\bibinfo
  {journal} {Nature Phys.}\ }\textbf {\bibinfo {volume} {16}},\ \bibinfo
  {pages} {132–142} (\bibinfo {year} {2020})}\BibitemShut {NoStop}%
\bibitem [{\citenamefont {Tada}\ \emph {et~al.}(2006)\citenamefont {Tada},
  \citenamefont {Kishimoto}, \citenamefont {Kominato}, \citenamefont {Shibata},
  \citenamefont {Yamada}, \citenamefont {Haseyama}, \citenamefont {Ogawa},
  \citenamefont {Funahashi}, \citenamefont {Yamamoto},\ and\ \citenamefont
  {Matsuki}}]{TADA2006488}%
  \BibitemOpen
  \bibfield  {author} {\bibinfo {author} {\bibfnamefont {M.}~\bibnamefont
  {Tada}}, \bibinfo {author} {\bibfnamefont {Y.}~\bibnamefont {Kishimoto}},
  \bibinfo {author} {\bibfnamefont {K.}~\bibnamefont {Kominato}}, \bibinfo
  {author} {\bibfnamefont {M.}~\bibnamefont {Shibata}}, \bibinfo {author}
  {\bibfnamefont {S.}~\bibnamefont {Yamada}}, \bibinfo {author} {\bibfnamefont
  {T.}~\bibnamefont {Haseyama}}, \bibinfo {author} {\bibfnamefont
  {I.}~\bibnamefont {Ogawa}}, \bibinfo {author} {\bibfnamefont
  {H.}~\bibnamefont {Funahashi}}, \bibinfo {author} {\bibfnamefont
  {K.}~\bibnamefont {Yamamoto}},\ and\ \bibinfo {author} {\bibfnamefont
  {S.}~\bibnamefont {Matsuki}},\ }\bibfield  {title} {\bibinfo {title}
  {Single-photon detection of microwave blackbody radiations in a
  low-temperature resonant-cavity with high {Rydberg} atoms},\ }\href
  {https://doi.org/https://doi.org/10.1016/j.physleta.2005.09.066} {\bibfield
  {journal} {\bibinfo  {journal} {Phys. Lett. A}\ }\textbf {\bibinfo {volume}
  {349}},\ \bibinfo {pages} {488} (\bibinfo {year} {2006})}\BibitemShut
  {NoStop}%
\bibitem [{\citenamefont {Fan}\ \emph {et~al.}(2014)\citenamefont {Fan},
  \citenamefont {Kumar}, \citenamefont {Daschner}, \citenamefont {K\"{u}bler},\
  and\ \citenamefont {Shaffer}}]{Fan:14}%
  \BibitemOpen
  \bibfield  {author} {\bibinfo {author} {\bibfnamefont {H.~Q.}\ \bibnamefont
  {Fan}}, \bibinfo {author} {\bibfnamefont {S.}~\bibnamefont {Kumar}}, \bibinfo
  {author} {\bibfnamefont {R.}~\bibnamefont {Daschner}}, \bibinfo {author}
  {\bibfnamefont {H.}~\bibnamefont {K\"{u}bler}},\ and\ \bibinfo {author}
  {\bibfnamefont {J.~P.}\ \bibnamefont {Shaffer}},\ }\bibfield  {title}
  {\bibinfo {title} {Subwavelength microwave electric-field imaging using
  {Rydberg} atoms inside atomic vapor cells},\ }\href
  {https://doi.org/10.1364/OL.39.003030} {\bibfield  {journal} {\bibinfo
  {journal} {Opt. Lett.}\ }\textbf {\bibinfo {volume} {39}},\ \bibinfo {pages}
  {3030} (\bibinfo {year} {2014})}\BibitemShut {NoStop}%
\bibitem [{\citenamefont {Sun}\ \emph {et~al.}(2017)\citenamefont {Sun},
  \citenamefont {Ma}, \citenamefont {Bai}, \citenamefont {Huang}, \citenamefont
  {Gao},\ and\ \citenamefont {Hou}}]{doi:10.1063/1.4997302}%
  \BibitemOpen
  \bibfield  {author} {\bibinfo {author} {\bibfnamefont {F.}~\bibnamefont
  {Sun}}, \bibinfo {author} {\bibfnamefont {J.}~\bibnamefont {Ma}}, \bibinfo
  {author} {\bibfnamefont {Q.}~\bibnamefont {Bai}}, \bibinfo {author}
  {\bibfnamefont {X.}~\bibnamefont {Huang}}, \bibinfo {author} {\bibfnamefont
  {B.}~\bibnamefont {Gao}},\ and\ \bibinfo {author} {\bibfnamefont
  {D.}~\bibnamefont {Hou}},\ }\bibfield  {title} {\bibinfo {title} {Measuring
  microwave cavity response using atomic rabi resonances},\ }\href@noop {}
  {\bibfield  {journal} {\bibinfo  {journal} {App. Phys. Lett.}\ }\textbf
  {\bibinfo {volume} {111}},\ \bibinfo {pages} {051103} (\bibinfo {year}
  {2017})}\BibitemShut {NoStop}%
\bibitem [{\citenamefont {Meyer}\ \emph {et~al.}(2021)\citenamefont {Meyer},
  \citenamefont {Kunz},\ and\ \citenamefont {Cox}}]{PhysRevApplied.15.014053}%
  \BibitemOpen
  \bibfield  {author} {\bibinfo {author} {\bibfnamefont {D.~H.}\ \bibnamefont
  {Meyer}}, \bibinfo {author} {\bibfnamefont {P.~D.}\ \bibnamefont {Kunz}},\
  and\ \bibinfo {author} {\bibfnamefont {K.~C.}\ \bibnamefont {Cox}},\
  }\bibfield  {title} {\bibinfo {title} {Waveguide-coupled {Rydberg} spectrum
  analyzer from 0 to 20 {G}hz},\ }\href
  {https://doi.org/10.1103/PhysRevApplied.15.014053} {\bibfield  {journal}
  {\bibinfo  {journal} {Phys. Rev. Appl.}\ }\textbf {\bibinfo {volume} {15}},\
  \bibinfo {pages} {014053} (\bibinfo {year} {2021})}\BibitemShut {NoStop}%
\bibitem [{\citenamefont {Gallagher}\ and\ \citenamefont
  {Cooke}(1979)}]{PhysRevLett.42.835}%
  \BibitemOpen
  \bibfield  {author} {\bibinfo {author} {\bibfnamefont {T.~F.}\ \bibnamefont
  {Gallagher}}\ and\ \bibinfo {author} {\bibfnamefont {W.~E.}\ \bibnamefont
  {Cooke}},\ }\bibfield  {title} {\bibinfo {title} {Interactions of blackbody
  radiation with atoms},\ }\href {https://doi.org/10.1103/PhysRevLett.42.835}
  {\bibfield  {journal} {\bibinfo  {journal} {Phys. Rev. Lett.}\ }\textbf
  {\bibinfo {volume} {42}},\ \bibinfo {pages} {835} (\bibinfo {year}
  {1979})}\BibitemShut {NoStop}%
\bibitem [{\citenamefont {Cooke}\ and\ \citenamefont
  {Gallagher}(1980)}]{PhysRevA.21.588}%
  \BibitemOpen
  \bibfield  {author} {\bibinfo {author} {\bibfnamefont {W.~E.}\ \bibnamefont
  {Cooke}}\ and\ \bibinfo {author} {\bibfnamefont {T.~F.}\ \bibnamefont
  {Gallagher}},\ }\bibfield  {title} {\bibinfo {title} {Effects of blackbody
  radiation on highly excited atoms},\ }\href
  {https://doi.org/10.1103/PhysRevA.21.588} {\bibfield  {journal} {\bibinfo
  {journal} {Phys. Rev. A}\ }\textbf {\bibinfo {volume} {21}},\ \bibinfo
  {pages} {588} (\bibinfo {year} {1980})}\BibitemShut {NoStop}%
\bibitem [{\citenamefont {Farley}\ and\ \citenamefont
  {Wing}(1981)}]{PhysRevA.23.2397}%
  \BibitemOpen
  \bibfield  {author} {\bibinfo {author} {\bibfnamefont {J.~W.}\ \bibnamefont
  {Farley}}\ and\ \bibinfo {author} {\bibfnamefont {W.~H.}\ \bibnamefont
  {Wing}},\ }\bibfield  {title} {\bibinfo {title} {Accurate calculation of
  dynamic stark shifts and depopulation rates of {Rydberg} energy levels
  induced by blackbody radiation. {H}ydrogen, helium, and alkali-metal atoms},\
  }\href {https://doi.org/10.1103/PhysRevA.23.2397} {\bibfield  {journal}
  {\bibinfo  {journal} {Phys. Rev. A}\ }\textbf {\bibinfo {volume} {23}},\
  \bibinfo {pages} {2397} (\bibinfo {year} {1981})}\BibitemShut {NoStop}%
\bibitem [{\citenamefont {Galvez}\ \emph {et~al.}(1995)\citenamefont {Galvez},
  \citenamefont {Lewis}, \citenamefont {Chaudhuri}, \citenamefont {Rasweiler},
  \citenamefont {Latvakoski}, \citenamefont {De~Zela}, \citenamefont
  {Massoni},\ and\ \citenamefont {Castillo}}]{PhysRevA.51.4010}%
  \BibitemOpen
  \bibfield  {author} {\bibinfo {author} {\bibfnamefont {E.~J.}\ \bibnamefont
  {Galvez}}, \bibinfo {author} {\bibfnamefont {J.~R.}\ \bibnamefont {Lewis}},
  \bibinfo {author} {\bibfnamefont {B.}~\bibnamefont {Chaudhuri}}, \bibinfo
  {author} {\bibfnamefont {J.~J.}\ \bibnamefont {Rasweiler}}, \bibinfo {author}
  {\bibfnamefont {H.}~\bibnamefont {Latvakoski}}, \bibinfo {author}
  {\bibfnamefont {F.}~\bibnamefont {De~Zela}}, \bibinfo {author} {\bibfnamefont
  {E.}~\bibnamefont {Massoni}},\ and\ \bibinfo {author} {\bibfnamefont
  {H.}~\bibnamefont {Castillo}},\ }\bibfield  {title} {\bibinfo {title}
  {Multistep transitions between {Rydberg} states of {Na} induced by blackbody
  radiation},\ }\href {https://doi.org/10.1103/PhysRevA.51.4010} {\bibfield
  {journal} {\bibinfo  {journal} {Phys. Rev. A}\ }\textbf {\bibinfo {volume}
  {51}},\ \bibinfo {pages} {4010} (\bibinfo {year} {1995})}\BibitemShut
  {NoStop}%
\bibitem [{\citenamefont {Beterov}\ \emph
  {et~al.}(2009{\natexlab{a}})\citenamefont {Beterov}, \citenamefont
  {Ryabtsev}, \citenamefont {Tretyakov},\ and\ \citenamefont
  {Entin}}]{PhysRevA.79.052504}%
  \BibitemOpen
  \bibfield  {author} {\bibinfo {author} {\bibfnamefont {I.~I.}\ \bibnamefont
  {Beterov}}, \bibinfo {author} {\bibfnamefont {I.~I.}\ \bibnamefont
  {Ryabtsev}}, \bibinfo {author} {\bibfnamefont {D.~B.}\ \bibnamefont
  {Tretyakov}},\ and\ \bibinfo {author} {\bibfnamefont {V.~M.}\ \bibnamefont
  {Entin}},\ }\bibfield  {title} {\bibinfo {title} {Quasiclassical calculations
  of blackbody-radiation-induced depopulation rates and effective lifetimes of
  {Rydberg} $ns$, $np$, and $nd$ alkali-metal atoms with
  $n\ensuremath{\le}80$},\ }\href {https://doi.org/10.1103/PhysRevA.79.052504}
  {\bibfield  {journal} {\bibinfo  {journal} {Phys. Rev. A}\ }\textbf {\bibinfo
  {volume} {79}},\ \bibinfo {pages} {052504} (\bibinfo {year}
  {2009}{\natexlab{a}})}\BibitemShut {NoStop}%
\bibitem [{\citenamefont {Beterov}\ \emph
  {et~al.}(2009{\natexlab{b}})\citenamefont {Beterov}, \citenamefont
  {Tretyakov}, \citenamefont {Ryabtsev}, \citenamefont {Entin}, \citenamefont
  {Ekers},\ and\ \citenamefont {Bezuglov}}]{Beterov_2009}%
  \BibitemOpen
  \bibfield  {author} {\bibinfo {author} {\bibfnamefont {I.~I.}\ \bibnamefont
  {Beterov}}, \bibinfo {author} {\bibfnamefont {D.~B.}\ \bibnamefont
  {Tretyakov}}, \bibinfo {author} {\bibfnamefont {I.~I.}\ \bibnamefont
  {Ryabtsev}}, \bibinfo {author} {\bibfnamefont {V.~M.}\ \bibnamefont {Entin}},
  \bibinfo {author} {\bibfnamefont {A.}~\bibnamefont {Ekers}},\ and\ \bibinfo
  {author} {\bibfnamefont {N.~N.}\ \bibnamefont {Bezuglov}},\ }\bibfield
  {title} {\bibinfo {title} {Ionization of {Rydberg} atoms by blackbody
  radiation},\ }\href {https://doi.org/10.1088/1367-2630/11/1/013052}
  {\bibfield  {journal} {\bibinfo  {journal} {New J. Phys.}\ }\textbf {\bibinfo
  {volume} {11}},\ \bibinfo {pages} {013052} (\bibinfo {year}
  {2009}{\natexlab{b}})}\BibitemShut {NoStop}%
\bibitem [{\citenamefont {Gallagher}\ \emph {et~al.}(1975)\citenamefont
  {Gallagher}, \citenamefont {Edelstein},\ and\ \citenamefont
  {Hill}}]{PhysRevA.11.1504}%
  \BibitemOpen
  \bibfield  {author} {\bibinfo {author} {\bibfnamefont {T.~F.}\ \bibnamefont
  {Gallagher}}, \bibinfo {author} {\bibfnamefont {S.~A.}\ \bibnamefont
  {Edelstein}},\ and\ \bibinfo {author} {\bibfnamefont {R.~M.}\ \bibnamefont
  {Hill}},\ }\bibfield  {title} {\bibinfo {title} {Radiative lifetimes of the
  $s$ and $d$ {Rydberg} levels of {Na}},\ }\href
  {https://doi.org/10.1103/PhysRevA.11.1504} {\bibfield  {journal} {\bibinfo
  {journal} {Phys. Rev. A}\ }\textbf {\bibinfo {volume} {11}},\ \bibinfo
  {pages} {1504} (\bibinfo {year} {1975})}\BibitemShut {NoStop}%
\bibitem [{\citenamefont {Kocher}\ and\ \citenamefont
  {Fairchild}(1978)}]{doi:10.1063/1.435914}%
  \BibitemOpen
  \bibfield  {author} {\bibinfo {author} {\bibfnamefont {C.~A.}\ \bibnamefont
  {Kocher}}\ and\ \bibinfo {author} {\bibfnamefont {C.~E.}\ \bibnamefont
  {Fairchild}},\ }\bibfield  {title} {\bibinfo {title} {Time‐of‐flight
  determination of radiative decay rates for high {Rydberg} states in atomic
  nitrogen},\ }\href@noop {} {\bibfield  {journal} {\bibinfo  {journal} {J.
  Chem. Phys.}\ }\textbf {\bibinfo {volume} {68}},\ \bibinfo {pages} {1884}
  (\bibinfo {year} {1978})}\BibitemShut {NoStop}%
\bibitem [{\citenamefont {Chupka}(1993)}]{doi:10.1063/1.465011}%
  \BibitemOpen
  \bibfield  {author} {\bibinfo {author} {\bibfnamefont {W.~A.}\ \bibnamefont
  {Chupka}},\ }\bibfield  {title} {\bibinfo {title} {Factors affecting
  lifetimes and resolution of {Rydberg} states observed in
  zero‐electron‐kinetic‐energy spectroscopy},\ }\href@noop {} {\bibfield
  {journal} {\bibinfo  {journal} {J. Chem. Phys.}\ }\textbf {\bibinfo {volume}
  {98}},\ \bibinfo {pages} {4520} (\bibinfo {year} {1993})}\BibitemShut
  {NoStop}%
\bibitem [{\citenamefont {Merkt}\ and\ \citenamefont
  {Zare}(1994)}]{doi:10.1063/1.467534}%
  \BibitemOpen
  \bibfield  {author} {\bibinfo {author} {\bibfnamefont {F.}~\bibnamefont
  {Merkt}}\ and\ \bibinfo {author} {\bibfnamefont {R.~N.}\ \bibnamefont
  {Zare}},\ }\bibfield  {title} {\bibinfo {title} {On the lifetimes of
  {Rydberg} states probed by delayed pulsed field ionization},\ }\href@noop {}
  {\bibfield  {journal} {\bibinfo  {journal} {J. Chem. Phys.}\ }\textbf
  {\bibinfo {volume} {101}},\ \bibinfo {pages} {3495} (\bibinfo {year}
  {1994})}\BibitemShut {NoStop}%
\bibitem [{\citenamefont {Oliveira}\ \emph {et~al.}(2002)\citenamefont
  {Oliveira}, \citenamefont {Mancini}, \citenamefont {Bagnato},\ and\
  \citenamefont {Marcassa}}]{PhysRevA.65.031401}%
  \BibitemOpen
  \bibfield  {author} {\bibinfo {author} {\bibfnamefont {A.~L.~d.}\
  \bibnamefont {Oliveira}}, \bibinfo {author} {\bibfnamefont {M.~W.}\
  \bibnamefont {Mancini}}, \bibinfo {author} {\bibfnamefont {V.~S.}\
  \bibnamefont {Bagnato}},\ and\ \bibinfo {author} {\bibfnamefont {L.~G.}\
  \bibnamefont {Marcassa}},\ }\bibfield  {title} {\bibinfo {title} {Measurement
  of {Rydberg}-state lifetimes using cold trapped atoms},\ }\href
  {https://doi.org/10.1103/PhysRevA.65.031401} {\bibfield  {journal} {\bibinfo
  {journal} {Phys. Rev. A}\ }\textbf {\bibinfo {volume} {65}},\ \bibinfo
  {pages} {031401} (\bibinfo {year} {2002})}\BibitemShut {NoStop}%
\bibitem [{\citenamefont {Feng}\ \emph {et~al.}(2011)\citenamefont {Feng},
  \citenamefont {Zhang}, \citenamefont {Zhang}, \citenamefont {Li},
  \citenamefont {Zhao},\ and\ \citenamefont {Jia}}]{fengzhigang}%
  \BibitemOpen
  \bibfield  {author} {\bibinfo {author} {\bibfnamefont {Z.-G.}\ \bibnamefont
  {Feng}}, \bibinfo {author} {\bibfnamefont {H.}~\bibnamefont {Zhang}},
  \bibinfo {author} {\bibfnamefont {L.-J.}\ \bibnamefont {Zhang}}, \bibinfo
  {author} {\bibfnamefont {C.-Y.}\ \bibnamefont {Li}}, \bibinfo {author}
  {\bibfnamefont {J.-M.}\ \bibnamefont {Zhao}},\ and\ \bibinfo {author}
  {\bibfnamefont {S.-T.}\ \bibnamefont {Jia}},\ }\bibfield  {title} {\bibinfo
  {title} {Measurement of lifetime of ultracold cesium {Rydberg} states},\
  }\href {https://doi.org/10.7498/aps.60.073202} {\bibfield  {journal}
  {\bibinfo  {journal} {Acta Phys. Sin.}\ }\textbf {\bibinfo {volume} {60}},\
  \bibinfo {pages} {073202} (\bibinfo {year} {2011})}\BibitemShut {NoStop}%
\bibitem [{\citenamefont {Mack}\ \emph {et~al.}(2015)\citenamefont {Mack},
  \citenamefont {Grimmel}, \citenamefont {Karlewski}, \citenamefont
  {S\'ark\'any}, \citenamefont {Hattermann},\ and\ \citenamefont
  {Fort\'agh}}]{PhysRevA.92.012517}%
  \BibitemOpen
  \bibfield  {author} {\bibinfo {author} {\bibfnamefont {M.}~\bibnamefont
  {Mack}}, \bibinfo {author} {\bibfnamefont {J.}~\bibnamefont {Grimmel}},
  \bibinfo {author} {\bibfnamefont {F.}~\bibnamefont {Karlewski}}, \bibinfo
  {author} {\bibfnamefont {L.}~\bibnamefont {S\'ark\'any}}, \bibinfo {author}
  {\bibfnamefont {H.}~\bibnamefont {Hattermann}},\ and\ \bibinfo {author}
  {\bibfnamefont {J.}~\bibnamefont {Fort\'agh}},\ }\bibfield  {title} {\bibinfo
  {title} {All-optical measurement of {Rydberg}-state lifetimes},\ }\href
  {https://doi.org/10.1103/PhysRevA.92.012517} {\bibfield  {journal} {\bibinfo
  {journal} {Phys. Rev. A}\ }\textbf {\bibinfo {volume} {92}},\ \bibinfo
  {pages} {012517} (\bibinfo {year} {2015})}\BibitemShut {NoStop}%
\bibitem [{\citenamefont {Gallagher}\ \emph {et~al.}(1976)\citenamefont
  {Gallagher}, \citenamefont {Humphrey}, \citenamefont {Hill},\ and\
  \citenamefont {Edelstein}}]{PhysRevLett.37.1465}%
  \BibitemOpen
  \bibfield  {author} {\bibinfo {author} {\bibfnamefont {T.~F.}\ \bibnamefont
  {Gallagher}}, \bibinfo {author} {\bibfnamefont {L.~M.}\ \bibnamefont
  {Humphrey}}, \bibinfo {author} {\bibfnamefont {R.~M.}\ \bibnamefont {Hill}},\
  and\ \bibinfo {author} {\bibfnamefont {S.~A.}\ \bibnamefont {Edelstein}},\
  }\bibfield  {title} {\bibinfo {title} {Resolution of $|{m}_{l}|$ and
  $|{m}_{j}|$ levels in the electric field ionization of highly excited $d$
  states of {Na}},\ }\href {https://doi.org/10.1103/PhysRevLett.37.1465}
  {\bibfield  {journal} {\bibinfo  {journal} {Phys. Rev. Lett.}\ }\textbf
  {\bibinfo {volume} {37}},\ \bibinfo {pages} {1465} (\bibinfo {year}
  {1976})}\BibitemShut {NoStop}%
\bibitem [{\citenamefont {Nosbaum}\ \emph {et~al.}(1995)\citenamefont
  {Nosbaum}, \citenamefont {Bleton}, \citenamefont {Cabaret}, \citenamefont
  {Yu}, \citenamefont {Gallagher},\ and\ \citenamefont
  {Pillet}}]{Nosbaum_1995}%
  \BibitemOpen
  \bibfield  {author} {\bibinfo {author} {\bibfnamefont {P.}~\bibnamefont
  {Nosbaum}}, \bibinfo {author} {\bibfnamefont {A.}~\bibnamefont {Bleton}},
  \bibinfo {author} {\bibfnamefont {L.}~\bibnamefont {Cabaret}}, \bibinfo
  {author} {\bibfnamefont {J.}~\bibnamefont {Yu}}, \bibinfo {author}
  {\bibfnamefont {T.~F.}\ \bibnamefont {Gallagher}},\ and\ \bibinfo {author}
  {\bibfnamefont {P.}~\bibnamefont {Pillet}},\ }\bibfield  {title} {\bibinfo
  {title} {Anticrossing spectroscopy of {Cs} {Rydberg} states},\ }\href
  {https://doi.org/10.1088/0953-4075/28/9/009} {\bibfield  {journal} {\bibinfo
  {journal} {J. Phys. B - At. Mol. Opt.}\ }\textbf {\bibinfo {volume} {28}},\
  \bibinfo {pages} {1707} (\bibinfo {year} {1995})}\BibitemShut {NoStop}%
\bibitem [{\citenamefont {Hollenstein}\ \emph {et~al.}(2001)\citenamefont
  {Hollenstein}, \citenamefont {Seiler}, \citenamefont {Schmutz}, \citenamefont
  {Andrist},\ and\ \citenamefont {Merkt}}]{doi:10.1063/1.1396856}%
  \BibitemOpen
  \bibfield  {author} {\bibinfo {author} {\bibfnamefont {U.}~\bibnamefont
  {Hollenstein}}, \bibinfo {author} {\bibfnamefont {R.}~\bibnamefont {Seiler}},
  \bibinfo {author} {\bibfnamefont {H.}~\bibnamefont {Schmutz}}, \bibinfo
  {author} {\bibfnamefont {M.}~\bibnamefont {Andrist}},\ and\ \bibinfo {author}
  {\bibfnamefont {F.}~\bibnamefont {Merkt}},\ }\bibfield  {title} {\bibinfo
  {title} {Selective field ionization of high {Rydberg} states: Application to
  zero-kinetic-energy photoelectron spectroscopy},\ }\href@noop {} {\bibfield
  {journal} {\bibinfo  {journal} {J. Chem. Phys.}\ }\textbf {\bibinfo {volume}
  {115}},\ \bibinfo {pages} {5461} (\bibinfo {year} {2001})}\BibitemShut
  {NoStop}%
\bibitem [{\citenamefont {Day}\ \emph {et~al.}(2008)\citenamefont {Day},
  \citenamefont {Brekke},\ and\ \citenamefont {Walker}}]{PhysRevA.77.052712}%
  \BibitemOpen
  \bibfield  {author} {\bibinfo {author} {\bibfnamefont {J.~O.}\ \bibnamefont
  {Day}}, \bibinfo {author} {\bibfnamefont {E.}~\bibnamefont {Brekke}},\ and\
  \bibinfo {author} {\bibfnamefont {T.~G.}\ \bibnamefont {Walker}},\ }\bibfield
   {title} {\bibinfo {title} {Dynamics of low-density ultracold {Rydberg}
  gases},\ }\href {https://doi.org/10.1103/PhysRevA.77.052712} {\bibfield
  {journal} {\bibinfo  {journal} {Phys. Rev. A}\ }\textbf {\bibinfo {volume}
  {77}},\ \bibinfo {pages} {052712} (\bibinfo {year} {2008})}\BibitemShut
  {NoStop}%
\bibitem [{\citenamefont {Branden}\ \emph {et~al.}(2009)\citenamefont
  {Branden}, \citenamefont {Juhasz}, \citenamefont {Mahlokozera}, \citenamefont
  {Vesa}, \citenamefont {Wilson}, \citenamefont {Zheng}, \citenamefont
  {Kortyna},\ and\ \citenamefont {Tate}}]{Branden_2009}%
  \BibitemOpen
  \bibfield  {author} {\bibinfo {author} {\bibfnamefont {D.~B.}\ \bibnamefont
  {Branden}}, \bibinfo {author} {\bibfnamefont {T.}~\bibnamefont {Juhasz}},
  \bibinfo {author} {\bibfnamefont {T.}~\bibnamefont {Mahlokozera}}, \bibinfo
  {author} {\bibfnamefont {C.}~\bibnamefont {Vesa}}, \bibinfo {author}
  {\bibfnamefont {R.~O.}\ \bibnamefont {Wilson}}, \bibinfo {author}
  {\bibfnamefont {M.}~\bibnamefont {Zheng}}, \bibinfo {author} {\bibfnamefont
  {A.}~\bibnamefont {Kortyna}},\ and\ \bibinfo {author} {\bibfnamefont {D.~A.}\
  \bibnamefont {Tate}},\ }\bibfield  {title} {\bibinfo {title} {Radiative
  lifetime measurements of rubidium {Rydberg} states},\ }\href
  {https://doi.org/10.1088/0953-4075/43/1/015002} {\bibfield  {journal}
  {\bibinfo  {journal} {J. Phys. B - At. Mol. Opt.}\ }\textbf {\bibinfo
  {volume} {43}},\ \bibinfo {pages} {015002} (\bibinfo {year}
  {2009})}\BibitemShut {NoStop}%
\bibitem [{\citenamefont {Fang}\ \emph {et~al.}(2001)\citenamefont {Fang},
  \citenamefont {Xie}, \citenamefont {Zhang}, \citenamefont {Hu},\ and\
  \citenamefont {Liu}}]{FANG2001469}%
  \BibitemOpen
  \bibfield  {author} {\bibinfo {author} {\bibfnamefont {D.-W.}\ \bibnamefont
  {Fang}}, \bibinfo {author} {\bibfnamefont {W.-J.}\ \bibnamefont {Xie}},
  \bibinfo {author} {\bibfnamefont {Y.}~\bibnamefont {Zhang}}, \bibinfo
  {author} {\bibfnamefont {X.}~\bibnamefont {Hu}},\ and\ \bibinfo {author}
  {\bibfnamefont {Y.-Y.}\ \bibnamefont {Liu}},\ }\bibfield  {title} {\bibinfo
  {title} {Radiative lifetimes of {Rydberg} state of ytterbium},\ }\href
  {https://doi.org/https://doi.org/10.1016/S0022-4073(00)00096-0} {\bibfield
  {journal} {\bibinfo  {journal} {J. Quant. Spectrosc. Radiat. Transf.}\
  }\textbf {\bibinfo {volume} {69}},\ \bibinfo {pages} {469} (\bibinfo {year}
  {2001})}\BibitemShut {NoStop}%
\bibitem [{\citenamefont {Du}\ \emph {et~al.}(2021)\citenamefont {Du},
  \citenamefont {Gong}, \citenamefont {Ji}, \citenamefont {Wang}, \citenamefont
  {Zhao}, \citenamefont {Xiao},\ and\ \citenamefont {Jia}}]{Du:21}%
  \BibitemOpen
  \bibfield  {author} {\bibinfo {author} {\bibfnamefont {J.}~\bibnamefont
  {Du}}, \bibinfo {author} {\bibfnamefont {T.}~\bibnamefont {Gong}}, \bibinfo
  {author} {\bibfnamefont {Z.}~\bibnamefont {Ji}}, \bibinfo {author}
  {\bibfnamefont {C.}~\bibnamefont {Wang}}, \bibinfo {author} {\bibfnamefont
  {Y.}~\bibnamefont {Zhao}}, \bibinfo {author} {\bibfnamefont {L.}~\bibnamefont
  {Xiao}},\ and\ \bibinfo {author} {\bibfnamefont {S.}~\bibnamefont {Jia}},\
  }\bibfield  {title} {\bibinfo {title} {Radiative lifetime measurement of
  ultracold cesium rydberg states by a simplified optical pumping method},\
  }\href {https://doi.org/10.1364/AO.411240} {\bibfield  {journal} {\bibinfo
  {journal} {Appl. Opt.}\ }\textbf {\bibinfo {volume} {60}},\ \bibinfo {pages}
  {276} (\bibinfo {year} {2021})}\BibitemShut {NoStop}%
\bibitem [{\citenamefont {Baltes}\ and\ \citenamefont {Hilf}(1976)}]{baltes}%
  \BibitemOpen
  \bibfield  {author} {\bibinfo {author} {\bibfnamefont {H.~P.}\ \bibnamefont
  {Baltes}}\ and\ \bibinfo {author} {\bibfnamefont {E.~R.}\ \bibnamefont
  {Hilf}},\ }\bibfield  {title} {\bibinfo {title} {Spectra of finite systems :
  a review of weyl's problem, the eigenvalue distribution of the wave equation
  for finite domains and its applications on the physics of small systems}\
  }(\bibinfo  {publisher} {Bibliographisches Institut,
  Mannheim/Wien/Z\"{u}rich},\ \bibinfo {year} {1976})\BibitemShut {NoStop}%
\bibitem [{\citenamefont {Sokolsky}\ and\ \citenamefont
  {Gorlach}(2014)}]{PhysRevA.89.013847}%
  \BibitemOpen
  \bibfield  {author} {\bibinfo {author} {\bibfnamefont {A.~A.}\ \bibnamefont
  {Sokolsky}}\ and\ \bibinfo {author} {\bibfnamefont {M.~A.}\ \bibnamefont
  {Gorlach}},\ }\bibfield  {title} {\bibinfo {title} {Scaling laws, pressure
  anisotropy, and thermodynamic effects for blackbody radiation in a finite
  cavity},\ }\href {https://doi.org/10.1103/PhysRevA.89.013847} {\bibfield
  {journal} {\bibinfo  {journal} {Phys. Rev. A}\ }\textbf {\bibinfo {volume}
  {89}},\ \bibinfo {pages} {013847} (\bibinfo {year} {2014})}\BibitemShut
  {NoStop}%
\bibitem [{\citenamefont {Reiser}\ and\ \citenamefont
  {Sch\"achter}(2013)}]{PhysRevA.87.033801}%
  \BibitemOpen
  \bibfield  {author} {\bibinfo {author} {\bibfnamefont {A.}~\bibnamefont
  {Reiser}}\ and\ \bibinfo {author} {\bibfnamefont {L.}~\bibnamefont
  {Sch\"achter}},\ }\bibfield  {title} {\bibinfo {title} {Geometric effects on
  blackbody radiation},\ }\href {https://doi.org/10.1103/PhysRevA.87.033801}
  {\bibfield  {journal} {\bibinfo  {journal} {Phys. Rev. A}\ }\textbf {\bibinfo
  {volume} {87}},\ \bibinfo {pages} {033801} (\bibinfo {year}
  {2013})}\BibitemShut {NoStop}%
\bibitem [{\citenamefont {Garc\'{\i}a-Garc\'{\i}a}(2008)}]{PhysRevA.78.023806}%
  \BibitemOpen
  \bibfield  {author} {\bibinfo {author} {\bibfnamefont {A.~M.}\ \bibnamefont
  {Garc\'{\i}a-Garc\'{\i}a}},\ }\bibfield  {title} {\bibinfo {title}
  {Finite-size corrections to the blackbody radiation laws},\ }\href
  {https://doi.org/10.1103/PhysRevA.78.023806} {\bibfield  {journal} {\bibinfo
  {journal} {Phys. Rev. A}\ }\textbf {\bibinfo {volume} {78}},\ \bibinfo
  {pages} {023806} (\bibinfo {year} {2008})}\BibitemShut {NoStop}%
\bibitem [{\citenamefont {Case}\ and\ \citenamefont
  {Chiu}(1970)}]{PhysRevA.1.1170}%
  \BibitemOpen
  \bibfield  {author} {\bibinfo {author} {\bibfnamefont {K.~M.}\ \bibnamefont
  {Case}}\ and\ \bibinfo {author} {\bibfnamefont {S.~C.}\ \bibnamefont
  {Chiu}},\ }\bibfield  {title} {\bibinfo {title} {Electromagnetic fluctuations
  in a cavity},\ }\href {https://doi.org/10.1103/PhysRevA.1.1170} {\bibfield
  {journal} {\bibinfo  {journal} {Phys. Rev. A}\ }\textbf {\bibinfo {volume}
  {1}},\ \bibinfo {pages} {1170} (\bibinfo {year} {1970})}\BibitemShut
  {NoStop}%
\bibitem [{\citenamefont {Heetae}\ \emph {et~al.}(2010)\citenamefont {Heetae},
  \citenamefont {Sukjoo},\ and\ \citenamefont {Soonjae}}]{kim}%
  \BibitemOpen
  \bibfield  {author} {\bibinfo {author} {\bibfnamefont {K.}~\bibnamefont
  {Heetae}}, \bibinfo {author} {\bibfnamefont {Y.}~\bibnamefont {Sukjoo}},\
  and\ \bibinfo {author} {\bibfnamefont {Y.}~\bibnamefont {Soonjae}},\
  }\bibfield  {title} {\bibinfo {title} {Finite size effect of one-dimensional
  thermal radiation},\ }\href@noop {} {\bibfield  {journal} {\bibinfo
  {journal} {J. Korean Phys. Soc.}\ }\textbf {\bibinfo {volume} {56}},\
  \bibinfo {pages} {554} (\bibinfo {year} {2010})}\BibitemShut {NoStop}%
\bibitem [{\citenamefont {Haroche}\ and\ \citenamefont
  {Raimond}(1985)}]{HAROCHE1985347}%
  \BibitemOpen
  \bibfield  {author} {\bibinfo {author} {\bibfnamefont {S.}~\bibnamefont
  {Haroche}}\ and\ \bibinfo {author} {\bibfnamefont {J.}~\bibnamefont
  {Raimond}},\ }\bibfield  {title} {\bibinfo {title} {Radiative properties of
  {Rydberg} states in resonant cavities}\ }(\bibinfo  {publisher} {Academic
  Press},\ \bibinfo {year} {1985})\ pp.\ \bibinfo {pages}
  {347--411}\BibitemShut {NoStop}%
\bibitem [{\citenamefont {Magnani}\ \emph {et~al.}(2020)\citenamefont
  {Magnani}, \citenamefont {Mojica-Casique},\ and\ \citenamefont
  {Marcassa}}]{Magnani_2020}%
  \BibitemOpen
  \bibfield  {author} {\bibinfo {author} {\bibfnamefont {B.}~\bibnamefont
  {Magnani}}, \bibinfo {author} {\bibfnamefont {C.}~\bibnamefont
  {Mojica-Casique}},\ and\ \bibinfo {author} {\bibfnamefont {L.~G.}\
  \bibnamefont {Marcassa}},\ }\bibfield  {title} {\bibinfo {title} {Finite size
  cavity effect on {nS} rubidium {Rydberg} state lifetimes},\ }\href
  {https://doi.org/10.1088/1361-6455/ab6a34} {\bibfield  {journal} {\bibinfo
  {journal} {J. Phys. B - At. Mol. Opt.}\ }\textbf {\bibinfo {volume} {53}},\
  \bibinfo {pages} {064004} (\bibinfo {year} {2020})}\BibitemShut {NoStop}%
\bibitem [{\citenamefont {Viteau}\ \emph {et~al.}(2010)\citenamefont {Viteau},
  \citenamefont {Radogostowicz}, \citenamefont {Chotia}, \citenamefont {Bason},
  \citenamefont {Malossi}, \citenamefont {Fuso}, \citenamefont {Ciampini},
  \citenamefont {Morsch}, \citenamefont {Ryabtsev},\ and\ \citenamefont
  {Arimondo}}]{Viteau_2010}%
  \BibitemOpen
  \bibfield  {author} {\bibinfo {author} {\bibfnamefont {M.}~\bibnamefont
  {Viteau}}, \bibinfo {author} {\bibfnamefont {J.}~\bibnamefont
  {Radogostowicz}}, \bibinfo {author} {\bibfnamefont {A.}~\bibnamefont
  {Chotia}}, \bibinfo {author} {\bibfnamefont {M.~G.}\ \bibnamefont {Bason}},
  \bibinfo {author} {\bibfnamefont {N.}~\bibnamefont {Malossi}}, \bibinfo
  {author} {\bibfnamefont {F.}~\bibnamefont {Fuso}}, \bibinfo {author}
  {\bibfnamefont {D.}~\bibnamefont {Ciampini}}, \bibinfo {author}
  {\bibfnamefont {O.}~\bibnamefont {Morsch}}, \bibinfo {author} {\bibfnamefont
  {I.~I.}\ \bibnamefont {Ryabtsev}},\ and\ \bibinfo {author} {\bibfnamefont
  {E.}~\bibnamefont {Arimondo}},\ }\bibfield  {title} {\bibinfo {title} {Ion
  detection in the photoionization of a {Rb} {Bose}{\textendash}{Einstein}
  condensate},\ }\href {https://doi.org/10.1088/0953-4075/43/15/155301}
  {\bibfield  {journal} {\bibinfo  {journal} {J. Phys. B - At. Mol. Opt.}\
  }\textbf {\bibinfo {volume} {43}},\ \bibinfo {pages} {155301} (\bibinfo
  {year} {2010})}\BibitemShut {NoStop}%
\bibitem [{\citenamefont {Viteau}\ \emph {et~al.}(2011)\citenamefont {Viteau},
  \citenamefont {Radogostowicz}, \citenamefont {Bason}, \citenamefont
  {Malossi}, \citenamefont {Ciampini}, \citenamefont {Morsch},\ and\
  \citenamefont {Arimondo}}]{Viteau:11}%
  \BibitemOpen
  \bibfield  {author} {\bibinfo {author} {\bibfnamefont {M.}~\bibnamefont
  {Viteau}}, \bibinfo {author} {\bibfnamefont {J.}~\bibnamefont
  {Radogostowicz}}, \bibinfo {author} {\bibfnamefont {M.~G.}\ \bibnamefont
  {Bason}}, \bibinfo {author} {\bibfnamefont {N.}~\bibnamefont {Malossi}},
  \bibinfo {author} {\bibfnamefont {D.}~\bibnamefont {Ciampini}}, \bibinfo
  {author} {\bibfnamefont {O.}~\bibnamefont {Morsch}},\ and\ \bibinfo {author}
  {\bibfnamefont {E.}~\bibnamefont {Arimondo}},\ }\bibfield  {title} {\bibinfo
  {title} {{Rydberg} spectroscopy of a {Rb} {MOT} in the presence of applied or
  ion created electric fields},\ }\href {https://doi.org/10.1364/OE.19.006007}
  {\bibfield  {journal} {\bibinfo  {journal} {Opt. Express}\ }\textbf {\bibinfo
  {volume} {19}},\ \bibinfo {pages} {6007} (\bibinfo {year}
  {2011})}\BibitemShut {NoStop}%
\bibitem [{\citenamefont {Faoro}\ \emph {et~al.}(2016)\citenamefont {Faoro},
  \citenamefont {Simonelli}, \citenamefont {Archimi}, \citenamefont {Masella},
  \citenamefont {Valado}, \citenamefont {Arimondo}, \citenamefont {Mannella},
  \citenamefont {Ciampini},\ and\ \citenamefont {Morsch}}]{PhysRevA.93.030701}%
  \BibitemOpen
  \bibfield  {author} {\bibinfo {author} {\bibfnamefont {R.}~\bibnamefont
  {Faoro}}, \bibinfo {author} {\bibfnamefont {C.}~\bibnamefont {Simonelli}},
  \bibinfo {author} {\bibfnamefont {M.}~\bibnamefont {Archimi}}, \bibinfo
  {author} {\bibfnamefont {G.}~\bibnamefont {Masella}}, \bibinfo {author}
  {\bibfnamefont {M.~M.}\ \bibnamefont {Valado}}, \bibinfo {author}
  {\bibfnamefont {E.}~\bibnamefont {Arimondo}}, \bibinfo {author}
  {\bibfnamefont {R.}~\bibnamefont {Mannella}}, \bibinfo {author}
  {\bibfnamefont {D.}~\bibnamefont {Ciampini}},\ and\ \bibinfo {author}
  {\bibfnamefont {O.}~\bibnamefont {Morsch}},\ }\bibfield  {title} {\bibinfo
  {title} {{van der Waals} explosion of cold {Rydberg} clusters},\ }\href
  {https://doi.org/10.1103/PhysRevA.93.030701} {\bibfield  {journal} {\bibinfo
  {journal} {Phys. Rev. A}\ }\textbf {\bibinfo {volume} {93}},\ \bibinfo
  {pages} {030701} (\bibinfo {year} {2016})}\BibitemShut {NoStop}%
\bibitem [{\citenamefont {B\'eguin}\ \emph {et~al.}(2013)\citenamefont
  {B\'eguin}, \citenamefont {Vernier}, \citenamefont {Chicireanu},
  \citenamefont {Lahaye},\ and\ \citenamefont
  {Browaeys}}]{PhysRevLett.110.263201}%
  \BibitemOpen
  \bibfield  {author} {\bibinfo {author} {\bibfnamefont {L.}~\bibnamefont
  {B\'eguin}}, \bibinfo {author} {\bibfnamefont {A.}~\bibnamefont {Vernier}},
  \bibinfo {author} {\bibfnamefont {R.}~\bibnamefont {Chicireanu}}, \bibinfo
  {author} {\bibfnamefont {T.}~\bibnamefont {Lahaye}},\ and\ \bibinfo {author}
  {\bibfnamefont {A.}~\bibnamefont {Browaeys}},\ }\bibfield  {title} {\bibinfo
  {title} {Direct measurement of the {van der Waals} interaction between two
  {Rydberg} atoms},\ }\href {https://doi.org/10.1103/PhysRevLett.110.263201}
  {\bibfield  {journal} {\bibinfo  {journal} {Phys. Rev. Lett.}\ }\textbf
  {\bibinfo {volume} {110}},\ \bibinfo {pages} {263201} (\bibinfo {year}
  {2013})}\BibitemShut {NoStop}%
\bibitem [{\citenamefont {Šibalić}\ \emph {et~al.}(2017)\citenamefont
  {Šibalić}, \citenamefont {Pritchard}, \citenamefont {Adams},\ and\
  \citenamefont {Weatherill}}]{SIBALIC2017319}%
  \BibitemOpen
  \bibfield  {author} {\bibinfo {author} {\bibfnamefont {N.}~\bibnamefont
  {Šibalić}}, \bibinfo {author} {\bibfnamefont {J.}~\bibnamefont
  {Pritchard}}, \bibinfo {author} {\bibfnamefont {C.}~\bibnamefont {Adams}},\
  and\ \bibinfo {author} {\bibfnamefont {K.}~\bibnamefont {Weatherill}},\
  }\bibfield  {title} {\bibinfo {title} {Arc: An open-source library for
  calculating properties of alkali {Rydberg} atoms},\ }\href
  {https://doi.org/https://doi.org/10.1016/j.cpc.2017.06.015} {\bibfield
  {journal} {\bibinfo  {journal} {Comput. Phys. Commun.}\ }\textbf {\bibinfo
  {volume} {220}},\ \bibinfo {pages} {319} (\bibinfo {year}
  {2017})}\BibitemShut {NoStop}%
\bibitem [{\citenamefont {Robertson}\ \emph {et~al.}(2021)\citenamefont
  {Robertson}, \citenamefont {Šibalić}, \citenamefont {Potvliege},\ and\
  \citenamefont {Jones}}]{ROBERTSON2021107814}%
  \BibitemOpen
  \bibfield  {author} {\bibinfo {author} {\bibfnamefont {E.}~\bibnamefont
  {Robertson}}, \bibinfo {author} {\bibfnamefont {N.}~\bibnamefont
  {Šibalić}}, \bibinfo {author} {\bibfnamefont {R.}~\bibnamefont
  {Potvliege}},\ and\ \bibinfo {author} {\bibfnamefont {M.}~\bibnamefont
  {Jones}},\ }\bibfield  {title} {\bibinfo {title} {Arc 3.0: An expanded python
  toolbox for atomic physics calculations},\ }\href
  {https://doi.org/https://doi.org/10.1016/j.cpc.2020.107814} {\bibfield
  {journal} {\bibinfo  {journal} {Comput. Phys. Commun.}\ }\textbf {\bibinfo
  {volume} {261}},\ \bibinfo {pages} {107814} (\bibinfo {year}
  {2021})}\BibitemShut {NoStop}%
\bibitem [{\citenamefont {Birch}\ \emph {et~al.}(1975)\citenamefont {Birch},
  \citenamefont {Cook}, \citenamefont {Harding}, \citenamefont {Jones},\ and\
  \citenamefont {Price}}]{Birch_1975}%
  \BibitemOpen
  \bibfield  {author} {\bibinfo {author} {\bibfnamefont {J.~R.}\ \bibnamefont
  {Birch}}, \bibinfo {author} {\bibfnamefont {R.~J.}\ \bibnamefont {Cook}},
  \bibinfo {author} {\bibfnamefont {A.~F.}\ \bibnamefont {Harding}}, \bibinfo
  {author} {\bibfnamefont {R.~G.}\ \bibnamefont {Jones}},\ and\ \bibinfo
  {author} {\bibfnamefont {G.~D.}\ \bibnamefont {Price}},\ }\bibfield  {title}
  {\bibinfo {title} {The optical constants of ordinary glass from $0.29$ to
  $4000\,\mathrm{cm^{-1}}$},\ }\href
  {https://doi.org/10.1088/0022-3727/8/11/014} {\bibfield  {journal} {\bibinfo
  {journal} {J. Phys. D Appl. Phys.}\ }\textbf {\bibinfo {volume} {8}},\
  \bibinfo {pages} {1353} (\bibinfo {year} {1975})}\BibitemShut {NoStop}%
\bibitem [{\citenamefont {Norrgard}\ \emph {et~al.}(2021)\citenamefont
  {Norrgard}, \citenamefont {Eckel}, \citenamefont {Holloway},\ and\
  \citenamefont {Shirley}}]{Norrgard_2021}%
  \BibitemOpen
  \bibfield  {author} {\bibinfo {author} {\bibfnamefont {E.~B.}\ \bibnamefont
  {Norrgard}}, \bibinfo {author} {\bibfnamefont {S.~P.}\ \bibnamefont {Eckel}},
  \bibinfo {author} {\bibfnamefont {C.~L.}\ \bibnamefont {Holloway}},\ and\
  \bibinfo {author} {\bibfnamefont {E.~L.}\ \bibnamefont {Shirley}},\
  }\bibfield  {title} {\bibinfo {title} {Quantum blackbody thermometry},\
  }\href {https://doi.org/10.1088/1367-2630/abe8f5} {\bibfield  {journal}
  {\bibinfo  {journal} {New J. Phys.}\ }\textbf {\bibinfo {volume} {23}},\
  \bibinfo {pages} {033037} (\bibinfo {year} {2021})}\BibitemShut {NoStop}%
\bibitem [{\citenamefont {Ovsiannikov}\ \emph {et~al.}(2011)\citenamefont
  {Ovsiannikov}, \citenamefont {Derevianko},\ and\ \citenamefont
  {Gibble}}]{PhysRevLett.107.093003}%
  \BibitemOpen
  \bibfield  {author} {\bibinfo {author} {\bibfnamefont {V.~D.}\ \bibnamefont
  {Ovsiannikov}}, \bibinfo {author} {\bibfnamefont {A.}~\bibnamefont
  {Derevianko}},\ and\ \bibinfo {author} {\bibfnamefont {K.}~\bibnamefont
  {Gibble}},\ }\bibfield  {title} {\bibinfo {title} {Rydberg spectroscopy in an
  optical lattice: Blackbody thermometry for atomic clocks},\ }\href
  {https://doi.org/10.1103/PhysRevLett.107.093003} {\bibfield  {journal}
  {\bibinfo  {journal} {Phys. Rev. Lett.}\ }\textbf {\bibinfo {volume} {107}},\
  \bibinfo {pages} {093003} (\bibinfo {year} {2011})}\BibitemShut {NoStop}%
\bibitem [{\citenamefont {Adams}\ \emph {et~al.}(2019)\citenamefont {Adams},
  \citenamefont {Pritchard},\ and\ \citenamefont {Shaffer}}]{Adams_2019}%
  \BibitemOpen
  \bibfield  {author} {\bibinfo {author} {\bibfnamefont {C.~S.}\ \bibnamefont
  {Adams}}, \bibinfo {author} {\bibfnamefont {J.~D.}\ \bibnamefont
  {Pritchard}},\ and\ \bibinfo {author} {\bibfnamefont {J.~P.}\ \bibnamefont
  {Shaffer}},\ }\bibfield  {title} {\bibinfo {title} {Rydberg atom quantum
  technologies},\ }\href {https://doi.org/10.1088/1361-6455/ab52ef} {\bibfield
  {journal} {\bibinfo  {journal} {J. Phys. B - At. Mol. Opt.}\ }\textbf
  {\bibinfo {volume} {53}},\ \bibinfo {pages} {012002} (\bibinfo {year}
  {2019})}\BibitemShut {NoStop}%
\end{thebibliography}%

\end{document}